\documentstyle[preprint,revtex]{aps}
\def\gtap{\raisebox{-.4ex}{\rlap{$\sim$}} \raisebox{.4ex}{$>$}}
\def\ltap{\raisebox{-.4ex}{\rlap{$\sim$}} \raisebox{.4ex}{$<$}}
\begin{document}
\thispagestyle{empty}
\font\fortssbx=cmssbx10 scaled \magstep2
\hbox{ 
\fortssbx University of Wisconsin Madison} 
\vspace{.3in}
\hfill\vbox{\hbox{\bf MAD/PH/711}
	    \hbox{September 1992}}\par
\vspace{.2in}
\begin{title}Supersymmetric Grand Unified Theories: Two Loop \\
Evolution of Gauge and Yukawa Couplings
\end{title}
\author{V.~Barger,  M.~S.~Berger, and
P.~Ohmann}
\begin{instit}
Physics Department, University of Wisconsin, Madison, WI 53706, USA\\
\end{instit}
\begin{abstract}
\baselineskip=18pt 
\nonum\section{abstract}
We make a numerical study of gauge and Yukawa
unification in supersymmetric grand unified models
and examine the quantitative
implications of fermion mass ans\"{a}tze at the grand unified
scale. Integrating the renormalization group equations with $\alpha _1(M_Z)$
and $\alpha _2(M_Z)$ as inputs, we find $\alpha _3(M_Z)\simeq 0.111 (0.122)$
for $M_{SUSY}^{}=m_t$ and
$\alpha _3(M_Z)\simeq 0.106 (0.116)$ for
$M_{SUSY}^{}=1$ TeV at one-loop (two-loop) order. Including
$b$ and $\tau $ Yukawa couplings in the evolution, we find an upper limit
$m_t\ltap 200$ GeV from Yukawa unification.
For given $m_t\ltap 175$ GeV, there are two solutions for $\beta$,
one with
$\tan \beta > m_t/m_b$, and one with
$\sin \beta \simeq 0.78(m_t/150\;{\rm GeV})$.
Taking a popular ansatz
for the mass matrices at the unified scale, we obtain a lower limit on the
top quark mass of $m_t\gtap 150 (115)$ GeV
for $\alpha _3(M_Z)=0.11 (0.12)$ and an upper limit on the supersymmetry
parameter $\tan \beta \ltap 50$ if $\alpha _3(M_Z)=0.11$.
The evolution of the quark mixing matrix elements is also evaluated.
\end{abstract}

\newpage
\section{Introduction}

There is renewed interest in supersymmetric grand unified theories (GUTs)
\cite{books} to explain gauge couplings, fermion masses and quark
mixings\cite{susygut,susygut2,Ramond1,Ramond3,DHR,Giudice,genmat,RR}. Recent
measurements of the gauge couplings at LEP and in other low energy
experiments\cite{Bethke,PDB}
are in reasonably good accord with expectations from minimal supersymmetric
GUTs with the scale of supersymmetry (SUSY) of order 1 TeV
or below\cite{susygut}.
Supersymmetric GUTs are also consistent with the non-observation to date of
proton decay\cite{protondecay}. In addition to the unification of gauge
couplings\cite{gut}, the unification of Yukawa couplings has been
considered to predict relations among quark masses\cite{CEG,EJ,Ramond2}.
With equal $b$-quark and $\tau$-lepton
Yukawa couplings at the GUT scale, the $m_b/m_{\tau}$ mass ratio is explained
by SUSY GUTs\cite{Ramond1,EJ}. With specific ans\"{a}tze for the
GUT scale mass matrices (e.g.
zero elements, mass hierarchy, relations of quark and lepton elements), other
predictions have been obtained from quark masses and mixings that are
consistent with measurements\cite{Ramond1,DHR,Giudice,KLN,BBHZ}. The
consideration
of fermion mass relationships has a long history\cite{Oakes,Fritzsch}
and includes single relations and mass matrices (``textures'') without
evolution\cite{GJ,noevol},
and single relations and mass matrices with evolution\cite{evol}.

Our approach is to explore supersymmetric GUTs first with the most general
assumptions, and then proceed to add additional GUT unification constraints to
obtain more predictions at the electroweak scale. The renormalization group
equations (RGEs)
used here are for the supersymmetric GUTs\cite{susyrge1,susyrge2}
with the minimal particle
content above the supersymmetry scale and the standard
model RGEs\cite{smrge} below the supersymmetry scale.
In \S II we explore the running of the gauge couplings in the
supersymmetric model at the two-loop level and compare the results to those
obtained at the one-loop level. Rather than try to predict the scale of
supersymmetry ($M_{SUSY}^{}$) which may be sensitive to unknown and model
dependent effects like particle thresholds at the GUT scale, we choose two
values of $M_{SUSY}^{}$ to illustrate the general trends that occur.
We also investigate the effects of the
Yukawa couplings on the gauge coupling running which enter at two
loops\cite{KLN} and have often been neglected in the past.
In \S III we explore the unification of Yukawa coupling constants.
First we consider the one-loop analytic solutions which can be obtained by
neglecting the bottom quark and tau Yukawa couplings $\lambda _b$ and
$\lambda _{\tau}$ relative to $\lambda _t$ in the RGEs. This serves as a useful
standard for comparison with the two-loop results for smaller values of
$\tan \beta $ ($<<m_t/m_b$), and many of the general features of the solutions
to the RGEs are already present at this stage.
We then investigate the two-loop RGE evolution of
the Yukawa couplings including the effects of $\lambda _b$,
$\lambda _{\tau}$, and $\lambda _t$.
Analytic solutions are not available for the two-loop
evolution, so we integrate the RGEs numerically.
In \S IV we investigate relations between Cabibbo-Kobayashi-Maskawa (CKM)
matrix elements and the ratios of quark masses. We investigate two popular
ans\"{a}tze\cite{DHR,Giudice,Ramond2} for Yukawa coupling matrices at the GUT
scale. Both of these ans\"{a}tze agree with all existing experimental data,
and this agreement is preserved at the two-loop level. We also integrate the
two-loop evolution equations for certain CKM matrix elements and quark mass
ratios in \S IV.
The two loop RGEs for both the minimal supersymmetric model and the standard
model are given in the appendix.

\section{Gauge Coupling Unification}

A consistent treatment to two loops in the running of the gauge couplings
involves the gauge couplings $g_i$ and the largest Yukawa couplings
$\lambda _t$, $\lambda _b$ and $\lambda _{\tau}$.
{}From general expressions\cite{susyrge1,susyrge2} that are summarized in the
appendix, we obtain the evolution equations
\begin{equation}
{{dg_i}\over {dt}}={g_i\over{16\pi^2 }}\left [b_ig_i^2+{1\over {16\pi^2 }}
\left (\sum _{j=1}^3b_{ij}g_i^2g_j^2-\sum _{j=t,b,\tau}a_{ij}g_i^2
\lambda _j^2\right )\right ] \label{dgidt} \;,
\end{equation}
%
%
%
%
\begin{eqnarray}
{{d\lambda _t}\over {dt}}={{\lambda _t}\over {16\pi ^2}}
\Bigg [\Bigg (&-&\sum c_ig_i^2+6\lambda _t^2+\lambda _b^2
\Bigg )
\nonumber \\
&+&{1\over {16\pi ^2}}\Bigg (\sum \left(c_ib_i+c_i^2/2\right )g_i^4
+g_1^2g_2^2+{136\over 45}g_1^2g_3^2+8g_2^2g_3^2\nonumber \\
&&\;\;\;\;\;\;+\lambda _t^2\left({6\over 5}g_1^2+6g_2^2+16g_3^2\right )+{2\over
5}
\lambda_b^2g_1^2 \nonumber \\
&&\;\;\;\;\;\;-\left \{22\lambda _t^4+5\lambda _t^2\lambda _b^2
+5\lambda _b^4+\lambda _b^2\lambda _{\tau}^2\right \}\Bigg )\Bigg ] \;,
\label{dytdt}
\end{eqnarray}
\begin{eqnarray}
{{d\lambda _b}\over {dt}}={{\lambda _b}\over {16\pi ^2}}
\Bigg [\Bigg (&-&\sum c_i^{\prime}g_i^2+\lambda _t^2
+6\lambda _b^2+\lambda _{\tau}^2
\Bigg )
\nonumber \\
&+&{1\over {16\pi ^2}}\Bigg (\sum \left(c_i^{\prime}b_i+c_i^{\prime 2}
/2\right )g_i^4
+g_1^2g_2^2+{8\over 9}g_1^2g_3^2+8g_2^2g_3^2\nonumber \\
&&\;\;\;\;\;\;+{4\over 5}\lambda_t^2g_1^2 +\lambda _b^2
\left({2\over 5}g_1^2+6g_2^2+16g_3^2\right )+{6\over 5}
\lambda_{\tau}^2g_1^2 \nonumber \\
&&\;\;\;\;\;\;-\left \{22\lambda _b^4+5\lambda _t^2\lambda _b^2
+3\lambda _b^2\lambda _{\tau}^2+3\lambda _{\tau}^4+5\lambda _t^4
\right \}\Bigg )\Bigg ] \;,
\label{dybdt}
\end{eqnarray}
\begin{eqnarray}
{{d\lambda _{\tau}}\over {dt}}={{\lambda _{\tau}}\over {16\pi ^2}}
\Bigg [\Bigg (&-&\sum c_i^{\prime \prime}g_i^2
+3\lambda _b^2+4\lambda _{\tau}^2
\Bigg )
\nonumber \\
&+&{1\over {16\pi ^2}}\Bigg (\sum \left(c_i^{\prime \prime}b_i
+c_i^{\prime \prime 2}/2\right )g_i^4
+{9\over 5}g_1^2g_2^2\nonumber \\
&&\;\;\;\;\;\;+\lambda _b^2\left(-{2\over 5}g_1^2+16g_3^2\right )
+\lambda_{\tau}^2\left({6\over 5}g_1^2+6g_2^2\right ) \nonumber \\
&&\;\;\;\;\;\;-\left \{3\lambda _t^2\lambda _b^2
+9\lambda _b^4+9\lambda _b^2\lambda _{\tau}^2+10\lambda _{\tau}^4
\right \}\Bigg )\Bigg ] \;,
\label{dytaudt}
\end{eqnarray}
The various coefficients in the above expressions are also
given in the appendix.
The variable is $t=\ln (\mu/M_G)$ where $\mu$ is the running mass
scale and $M_G$ is the GUT unification mass.
The renormalization group equations of dimensionless parameters like the gauge
couplings and Yukawa couplings are independent of the dimensionful
soft-supersymmetry breaking parameters.

We begin with the recent values of $\alpha _{em}$ and
$\sin ^2 \hat{\theta}_W^{}$
at scale $M_Z=91.17$ GeV given in the 1992 Particle Data Book\cite{PDB,Lang}
\begin{mathletters}
\begin{eqnarray}
(\alpha _{em})^{-1}&=&127.9\pm0.2\;, \\
\sin ^2 \hat{\theta}_W^{}&=&0.2326\pm0.0008\;,
\end{eqnarray}
\end{mathletters}
where $\hat{\theta}_W^{}$ refers to the weak angle in the modified
minimal subtraction $\overline{{\rm MS}}$
scheme\cite{Sirlin}. These values
correspond to electroweak gauge couplings of
\begin{mathletters}
\begin{eqnarray}
\alpha _1(M_Z)^{-1}&=&58.89\pm0.11\;, \\
\alpha _2(M_Z)^{-1}&=&29.75\pm0.11\;, \label{alphain}
\end{eqnarray}
\end{mathletters}
For simplicity we initially set the supersymmetric scale $M_{SUSY}^{}$ equal
to the top quark mass $m_t$ and set all Yukawa contributions in Eq.
(\ref{dgidt}) to zero. Then evolving $\alpha _1$ and $\alpha _2$ from
scale $M_Z$ up to scale $m_t$, we have
\begin{mathletters}
\begin{eqnarray}
\alpha _1(m_t)^{-1}&=&\alpha _1(M_Z)^{-1}+{{53}\over {30\pi }}\ln (M_Z/m_t)
\;, \\
\alpha _2(m_t)^{-1}&=&\alpha _2(M_Z)^{-1}-{{11}\over {6\pi }}\ln (M_Z/m_t) \;,
\label{alpha12}
\end{eqnarray}
\end{mathletters}
We use the value $M_Z=91.17$ GeV, neglecting its experimental uncertainty.

Next, for a grid of $\alpha _G^{}$ and $M_G^{}$ values, we evolve from
the GUT scale down to
the chosen $m_t$ scale and retain those GUT scale inputs for which Eqs.
(6) and (7) are satisfied. We use the two-loop SUSY GUT unification condition
$\alpha _G^{}=\alpha _1(M_G^{})=\alpha _2(M_G^{})$. For the acceptable
GUT inputs we also evolve the strong coupling $\alpha _3(M_G^{})=\alpha _G^{}$
down to scale $m_t$ and then use 3-loop
QCD to further evolve it to scale $M_Z$. The three-loop expression
\begin{equation}
\alpha _3(\mu)^{-1}=-{{b_0}\over 2}
\ln \left ({{\mu ^2}\over {\Lambda ^2}}\right )
+{{b_1}\over {b_0}}\ln\left (\ln {{\mu ^2}\over {\Lambda ^2}}\right )
-2{{b_1^2}\over {b_0^3}}\left [
\ln \left (\ln {{\mu ^2}\over {\Lambda ^2}}\right )-\left (
{{b_0b_2}\over {b_1^2}}-1\right )\right ]\left (
\ln {{\mu ^2}\over {\Lambda ^2}}\right )^{-1}\;,
\label{alpha3} \end{equation}
with the $b_i$ given in Ref.~\cite{Marciano}, is iteratively solved to find
$\Lambda$ from $\alpha _3(m_t)$. Eq.~(\ref{alpha3}) is then evaluated
for $\mu = M_Z$ to obtain $\alpha _3(M_Z)$. The resulting values for $\Lambda$
for two representative values of $\alpha _3(M_Z)$ are given in Table 1.

\newpage
{\center \begin{tabular}{|c|c|c|c|}
\hline
\multicolumn{1}{|c|}{$\alpha _3(M_Z)$}
&\multicolumn{1}{|c|}{$\Lambda ^{(5)}$}
&\multicolumn{1}{|c|}{$\Lambda ^{(4)}$}
&\multicolumn{1}{|c|}{$\Lambda ^{(3)}$}
\\ \hline \hline
\multicolumn{1}{|c|}{0.11}
&\multicolumn{1}{|c|}{129.1}
&\multicolumn{1}{|c|}{188.3}
&\multicolumn{1}{|c|}{225.0}
\\ \hline
\multicolumn{1}{|c|}{0.12}
&\multicolumn{1}{|c|}{233.4}
&\multicolumn{1}{|c|}{320.2}
&\multicolumn{1}{|c|}{360.0}
\\ \hline
\end{tabular}
\vskip .3in }
\begin{center}
{\bf Table 1:}  The QCD parameter $\Lambda ^{(n_f)}$ in MeV,\\
where $n_f$ is the number of active flavors.
\end{center}
\vskip .6in

We also investigate the effects of taking a supersymmetry scale
higher than $m_t$.
Below $M_{SUSY}^{}$, the RGE are similar to the
non-supersymmetric standard model. A linear combination of the Higgs doublets
is integrated out of the theory at $M_{SUSY}^{}$ leaving the orthogonal
combination
$\Phi_{(SM)}^{}=\Phi _d\cos \beta+\tilde{\Phi}_u\sin \beta$
coupled to the fermions in a way that depends on
$\tan \beta$\cite{Ramond1,GHS,HH}; this combination
results from the assumption that the
three soft-supersymmetry breaking parameters in the Higgs potential can be
equated to $M_{SUSY}^{}$. We use the two-loop RGEs\cite{smrge}
for the standard
model, matching the couplings at $M_{SUSY}^{}$.
Taking a single SUSY scale is an idealized situation since in general the
supersymmetric particle spectrum is spread over a range of masses\cite{RR}.
Without further assumptions we cannot predict this spectrum.
Given that such uncertainties exist, the
predicted range for $\alpha _3$ should be taken to
be representative only.

The ranges of $\alpha _G^{-1}$ and $M_G^{}$ parameters obtained from the
procedure outlined above are presented in Fig. 1 for one-loop and two-loop
evolution with the choices $M_{SUSY}^{}=m_t$ and $M_{SUSY}^{}=1$ TeV.
The shaded regions
denote the allowed GUT parameter space.
The two-loop values obtained for $\alpha _G^{}$ and $M_G^{}$ are higher
than the one-loop values and
consequently $\alpha _3(M_Z)$ is higher for the two-loop evolution.
Note that raising the SUSY scale from $m_t$ to 1 TeV lowers $M_G^{}$ and
$\alpha _G^{}$; hence $\alpha _3(M_Z)$ decreases as well.

Figure 2 shows the corresponding results of the two-loop evolution over the
full range of $\mu $. We find
the ranges for $\alpha _3(M_Z)$ with $m_t=150$ GeV shown in Table 2.
\vskip 0.5in
{\center \begin{tabular}{|c|c|c|}
\hline
\multicolumn{1}{|c|}{$M_{SUSY}^{}$}
&\multicolumn{1}{|c|}{one-loop}
&\multicolumn{1}{|c|}{two-loop}
\\ \hline \hline
\multicolumn{1}{|c|}{$m_t= 150$ GeV}
&\multicolumn{1}{|c|}{$0.1112\pm 0.0024$}
&\multicolumn{1}{|c|}{$0.1224\pm 0.0033$}
\\ \hline
\multicolumn{1}{|c|}{1 TeV}
&\multicolumn{1}{|c|}{$0.1065\pm 0.0024$}
&\multicolumn{1}{|c|}{$0.1161\pm 0.0028$}
\\ \hline
\end{tabular}
\vskip .3in }
\begin{center}
{\bf Table 2:}  Ranges obtained for $\alpha _3(M_Z)$ from the \\
input values $\alpha _{em}$ and $\sin ^2 \hat{\theta}_W^{}$.
\end{center}
\vskip .5in
The two-loop values of $\alpha _3(M_Z)$ are about $10\%$
larger than the one-loop values. The effect of the higher SUSY scale is to
lower $\alpha _3(M_Z)$ by about 5\%.

Inclusion of Yukawa couplings in the two-loop
evolution also lowers the value of $\alpha _3(M_Z)$ somewhat.
For example setting
$\lambda _t=\lambda _b=\lambda _{\tau}=1$ at the GUT scale, we obtain
a two-loop value of $\alpha _3(M_Z)=0.1189\pm0.0031$ for $M_{SUSY}^{}=m_t$.

The effects on the gauge couplings
of including the Yukawa couplings in the evolution are rather small for Yukawa
couplings in the perturbative regime, justifying their neglect in most
previous analyses; for large values of $\tan \beta$ the changes in the gauge
couplings due to inclusion of Yukawa couplings can be a few percent.

The experimental situation regarding the determination of $\alpha _3$ is
presently somewhat clouded\cite{Bethke},
with deep inelastic scattering determinations in the range
of the one-loop calculations in Table 2 and LEP determinations similar to the
two-loop results of Table 2.

There are other uncertainties not taken into account here,
due to threshold corrections from the
unknown particle content at the heavy scale\cite{Weinberg,Hall,barhall},
which can also change the $\alpha _3$ values obtained above.
These corrections
are model-dependent so we have not attempted to include such contributions.
However recent analysis have shown that the constraints from non-observation
of proton decay greatly reduce the potential uncertainties from GUT
thresholds\cite{KLN,threshold}.

\section{Yukawa Unification}

\noindent {\bf A. One-loop analytic results}

The unification of Yukawa couplings first introduced  by Chanowitz, Ellis and
Gaillard\cite{CEG} has been reconsidered recently
\cite{Ramond1,DHR,Giudice,KLN,GHS}.
The GUT scale condition $\lambda _b(M_G^{})=\lambda _{\tau }(M_G^{})$
leads to  a successful prediction for the mass ratio $m_b/m_{\tau}$
provided that a low energy supersymmetry exists\cite{Ramond1}.
The $b$ to $\tau$ mass ratio is given by
\begin{equation}
{{m_b}\over {m_{\tau}}}={{\eta_b}\over {\eta _{\tau}}}R_{b/\tau}(m_t)\;,
\end{equation}
where
\begin{equation}
R_{b/\tau}(m_t)\equiv {{\lambda _b(m_t)}\over {\lambda _{\tau}(m_t)}}
={{m_b(m_t)}\over {m_{\tau}(m_t)}}\;,
\end{equation}
is the $b$ to $\tau$ ratio of running masses at scale $m_t$ and
\begin{equation}
\eta_f={{m_f(m_f)}\over {m_f(m_t)}} \label{etadef} \;\;\;\;\;\;\;\;
{\rm if} \;\;\; m_f > 1 GeV \;,\\
\end{equation}
\begin{equation}
\eta_f={{m_f(1 {\rm GeV})}\over {m_f(m_t)}} \label{etadef2} \;\;\;\;\;\;\;\;
{\rm if} \;\;\; m_f < 1 GeV
\;,\\
\end{equation}
is a scaling factor including both QCD and QED effects in the running mass
below $m_t$. We have determined the $\eta _f$
scaling factors to three-loop order in QCD and
one-loop order in QED. The QCD running of the quark mass is described by
\begin{eqnarray}
m_q(\mu )&=&\hat{m_q}
\left (2b_0\alpha _3\right )^{\gamma _0/b_0}\Big [
1+\left ({\gamma _1\over b_0}-{{\gamma _0b_1}\over {b_0^2}}\right )\alpha _3
\nonumber \\
&&+{1\over 2}\left [\left ({\gamma _1\over b_0}-{{\gamma _0b_1}\over {b_0^2}}
\right )^2+\left ({\gamma _2\over b_0}+{{\gamma _0b_1^2}\over {b_0^3}}
-{{b_1\gamma _1+b_2\gamma _0}\over {b_0^2}}\right )\right ]\alpha _3^2
+{\cal O}(\alpha _3^3)\Big ]\;, \label{massrun}
\end{eqnarray}
where the anomalous dimensions $\gamma _0$, $\gamma _1$ and
$\gamma _2$ are given in Ref.~\cite{Gorishny}. The scale-invariant mass
$\hat{m_q}$ cancels in the ratio in Eq.~(\ref{etadef}).
The one-loop QED running from scale $\mu ^{\prime}$ to scale $\mu $ introduces
modifications
\begin{equation}
m_f(\mu)=m_f(\mu ^{\prime})
\left ({{\alpha (\mu )}\over {\alpha (\mu ^{\prime})}} \right )
^{\gamma _0^{QED}/b_0^{QED}}\;,
\end{equation}
where the QED beta function and anomalous dimension are
given by\cite{Hall}
\begin{eqnarray}
b_0^{QED}&=&{4\over 3}\left (3\sum Q_u^2+3\sum Q_d^2+\sum Q_e^2\right )\;, \\
\gamma _0^{QED}&=&-3Q_f^2\;,
\end{eqnarray}
and the sums run over the active fermions at the relevant scale.
The dependence of the QCD-QED scaling factors $\eta $ on $\alpha _3(M_Z)$ is
shown in Figure 3; these factors increase as $\alpha _3(M_Z)$ increases.

We note that the physical top mass is related to
the running mass by\cite{Tarrach}
\begin{equation}
m_t^{phys}=m_t(m_t)\left [1+{4\over {3\pi}}\alpha _3(m_t)
+{\cal O}(\alpha _3^2)\right ]\;.
\end{equation}

The effects of the top quark Yukawa $\lambda _t$ can be studied
semi-analytically at one-loop neglecting the effects of the bottom and tau
Yukawa couplings $\lambda _b$ and $\lambda _{\tau}$ in Eqs.~(\ref{dgidt}) and
(\ref{dytdt}), which is a valid approximation for small to moderate
$\tan \beta $ (i.e. $\tan \beta \ltap 10$). Following Ref.~\cite{DHR}
we find\cite{translate}
\begin{equation}
{{m_b}\over {m_{\tau}}}=y{{\eta ^{1/2}}\over {x}}{{\eta _b}\over
{\eta _{\tau}}}\;, \label{btau}
\end{equation}
where $x(\mu )$, $y(\mu )$, $\eta (\mu )$ defined by
\begin{eqnarray}
x(\mu ) &=& (\alpha_G^{}/\alpha_1(\mu))^{1/6} (\alpha_G^{}/\alpha_2(\mu))
^{3/2} \;, \\
y(\mu ) &=& \exp \left \{ -{1\over {16\pi ^2}}\int\limits_{\mu}^{M_G}
\lambda _t^2(\mu')d\ln\mu' \right \} \label{defy} \;, \\
\eta(\mu) &=& \prod_{i=1,2,3}(\alpha_G^{}/\alpha_i(\mu))^{c_i/b_i} \;,
\end{eqnarray}
are to be evaluated at $\mu =m_t$ in Eq.~(\ref{btau}).
Henceforth $x$, $y$, $\eta $ shall be understood as being evaluated at scale
$m_t$ when an argument is not explicitly specified.
Typical values of these quantities obtained in Ref.~\cite{BBHZ} are
$x=1.52$, $y=0.75-0.81$, $\eta =10.3$
for a bottom mass given by the Gasser-Leutwyler (GL) QCD sum rule
determination $m_b=4.25\pm0.1$ GeV\cite{GL}
taken within its 90\% confidence range and $\alpha _3=0.111$.
The quantity $y$ gives the scaling from $M_G$ to $m_t$ that
arises from a heavy quark,
beyond the scaling due to the gauge couplings.
The factor $y(m_t)$ is constrained
to lie in a narrow range of
values by Eq.~(\ref{btau}). The integral in Eq.~(\ref{defy}) is crucial in
explaining the $m_b/m_{\tau}$ ratio. In fact if $\lambda _t$ is neglected then
$y=1$ and the $m_b/m_{\tau }$ ratio is found to be too large.

For a given value of $m_t$, there exist two solutions for $\tan \beta$.
This fact can be understood qualitatively by studying the one-loop RGE for
$R_{b/\tau}\equiv \lambda _b/\lambda _{\tau}$.
\begin{equation}
{{dR_{b/\tau}}\over {dt}}={{R_{b/\tau}}\over {16\pi ^2}}
\left (-\sum d_ig_i^2+\lambda _t^2
+3\lambda _b^2-3\lambda _{\tau}^2\right )\;.
\label{dRdt}
\end{equation}
For small $\tan \beta$
the bottom and tau Yukawas do not play a significant role in the RGE, and any
particular value for $m_b/m_{\tau}$ is obtained for a unique value of
$\lambda _t(m_t)$,
which corresponds to a linear relationship between $m_t$ and $\sin
\beta$,
\begin{equation}
{{m_t}\over{\sin \beta}} ={v\over {\sqrt{2}}}\lambda _t(m_t)
= \pi v\sqrt{{2\eta }\over {3I}}
\left[ 1-y^{12} \right]^{1/2}\;, \label{linear}
\end{equation}
where $v = 246$ GeV and
\begin{equation}
I = \int\limits_{m_t}^{M_G^{}}\eta(\mu') d\ln\mu' \;. \label{defI}
\end{equation}
The numerical value for $I$ from Ref.~\cite{BBHZ} is 113.79 for $m_t=170$ GeV.
For large $\tan \beta$, where the effects of $\lambda _b$ and $\lambda _{\tau}$
on the running Yukawa couplings
can be substantial, an increase in $\lambda _b$ can be compensated in the RGE
by a
decrease in $\lambda _t$. Hence, for increasing $\tan \beta$, the correct
prediction for $m_b/m_{\tau}$ is obtained for decreased values of the top
quark Yukawa. Thus there is a second solution to the RGE for $R_{b/\tau}$
with a large value of $\tan \beta$.
The inclusion of the two-loop effects does not alter these observations.

The one-loop RGE for $R_{s/\mu}\equiv \lambda_s/\lambda _{\mu}$
\begin{equation}
{{dR_{s/\mu}}\over {dt}}={{R_{s/\mu}}\over {16\pi ^2}}
\left (-\sum d_ig_i^2\right )\;.
\label{dRpdt}
\end{equation}
is similar to
Eq.~(\ref{dRdt}), except that it receives no contribution from the dominant
Yukawa couplings $\lambda _t$, $\lambda _b$, and $\lambda _{\tau}$.
When the value $R_{s/\mu}(M_G)=1/3$ is assumed at the GUT scale, the
prediction at the electroweak scale is
\begin{equation}
{{m_s}\over {m_{\mu}}}={1\over 3}{{\eta ^{1/2}}\over {x}}
{{\eta _s}\over {\eta _{\mu}}} \;.
\end{equation}
Notice that this equation does not include the scaling parameter $y$ because
the top quark Yukawa does not affect the running of the second generation
quarks and leptons.
This relation for $m_s/m_{\mu}$
is in good agreement with the experimental values, but it is not
as stringent as the $m_b/m_{\tau}$ relation due to the sizable
uncertainty in the strange quark mass. The result $m_s/m_{\mu}= 1.54$ was
obtained in Ref.~\cite{BBHZ}, to be compared with the GL
determination\cite{GL} $m_s/m_{\mu}= 1.66\pm 0.52$.

A popular strategy is to relate the mixing angles in the
CKM matrix to
ratios of quark masses, taking into account the evolution from the GUT
scale in non-SUSY\cite{Ma} or SUSY\cite{KSBabu} models.
For example, one popular GUT scale ansatz is $|V_{cb}|\approx \sqrt{m_c/m_t}$
which requires a GUT boundary condition on
$R_{c/t}\equiv \lambda_c/\lambda _t$ of
\begin{equation}
\sqrt{R_{c/t}(M_G)}=|V_{cb}(M_G)|\;, \label{GUTbc}
\end{equation}
The one-loop SUSY RGE for
$R_{c/t}$ is
\begin{equation}
{{dR_{c/t}}\over {dt}}=-{{R_{c/t}}\over {16\pi ^2}}
\Bigg [3\lambda _t^2
+\lambda _b^2\Bigg] \;. \label{dRppdt1}
\end{equation}
The corresponding one-loop SUSY RGE for the running CKM matrix element
$|V_{cb}|$ is\cite{KSBabu},
\begin{equation}
{{d|V_{cb}|}\over {dt}}=-{{|V_{cb}|}\over {16\pi ^2}}
\Bigg [\lambda _t^2
+\lambda _b^2\Bigg ] \;, \label{dVcbdt1}
\end{equation}
The pure gauge coupling parts of the RGEs are not present in
Eqs.~(\ref{dRppdt1}) and (\ref{dVcbdt1}) since $R_{c/t}$ and $V_{cb}$
are ratios of
elements from the up quark Yukawa matrix and the down quark Yukawa matrix.

Neglecting the non-leading effects of $\lambda _b$,
the one-loop results of Ref.~\cite{DHR} at the electroweak scale obtained from
evolution are
\begin{equation}
|V_{cb}(m_t)|=|V_{cb}(M_G)|y^{-1}\quad \quad
R_{c/t}(m_t)=R_{c/t}(M_G)y^{-3}\;,
\end{equation}
or equivalently using Eq.~(\ref{GUTbc})
\begin{equation}
|V_{cb}(m_t)|=\sqrt{{{ym_c}\over{\eta _cm_t}}}\;. \label{Vcbsimple}
\end{equation}

Since $y$ is already well constrained by the $b$-mass relation of
Eq.~(\ref{btau}) (for the one-loop value of $\alpha _3(M_Z)=0.111$),
Eq.~(\ref{Vcbsimple})
requires that $m_t$ must be large in order that $|V_{cb}|$ falls in the
experimentally allowed range $0.032-0.054$ (and even then $|V_{cb}|$ is found
to be at the
upper limit of its allowed range). If, however, we use a larger
value of $\alpha _3(M_Z)$ indicated by the two-loop equations, say $0.12$,
then $\eta _c$ increases by about 14\%, as shown in Figure 3.
Furthermore the increased values of the
scaling parameters $\eta $ and $\eta _b$ require about a 9\% decrease in $y$
to explain the $m_b/m_{\tau}$ ratio
in Eq.~(\ref{btau}). The resulting
$|V_{cb}|$ is reduced by about 12\% and is then closer to its central
experimental value.
Of course, a consistent treatment at the two-loop level requires the two-loop
generalization of Eq.~(\ref{Vcbsimple}) obtained
by solving the full set of RGEs.
One of the questions we will address subsequently is for what values of
$m_t$ and $\tan \beta$ can the $|V_{cb}|$
and the $m_b/m_{\tau}$ constraints be realized simultaneously.

The predictions above are all based upon the assumption that the couplings
remain in the perturbative regime during the evolution from the GUT scale down
to the electroweak scale. Otherwise it is not valid to use the RGEs which are
calculated order by order in perturbation theory. One can impose
this perturbative unification condition as a constraint. For
$m_b$ at the lower end of the GL QCD sum rule range $4.1 - 4.4$ GeV
the top quark Yukawa coupling at the GUT scale,
$\lambda _t(M_G^{})$, becomes large, as can be demonstrated from analytic
solutions to the one-loop RGEs in the approximation that $\lambda _b$ and
$\lambda _{\tau}$ are neglected compared to $\lambda _t$
(valid for small to moderate $\tan \beta $).

The top quark Yukawa at the GUT scale is given by
\begin{equation}
\lambda _t(M_G^{})^2={{4\pi ^2}\over {3I}}\left [
{{1}\over {y ^{12}}}-1\right ]\;.
\label{ltG}
\end{equation}
Taking\cite{BBHZ}
$\alpha _3(M_Z)=0.111$ and $m_b=4.25$ GeV and $m_t=170$ GeV
gives $\lambda _t(M_G^{})=1.5$.
Larger values of $\alpha _3(M_Z)$ lead to increased $\eta _b$ via
Eq.~(\ref{etadef}) giving smaller $y$ in Eq.~(\ref{btau}) and a
correspondingly larger value of $\lambda _t(M_G^{})$.
The quantity $\lambda _t(M_G)$ is plotted versus $\alpha _3(M_Z)$ in Figure 4.
Larger values
of $\alpha _3(M_Z)\approx 0.12$ can yield
$\lambda _t(M_G^{})\gtap 3$ that cast the perturbative unification in doubt.
Keeping the gauge couplings
fixed and varying $m_b$, one sees that smaller values of $m_b$ also yield
larger values of $\lambda _t(M_G^{})$.

The scaling
parameter $y$ is manifestly less than one  by Eq.~(\ref{defy}) since
$\lambda _t^2>0$ in the region $m_t<\mu <M_G^{}$.
This implies an upper limit on
$m_b$ in Eq.~(\ref{btau}) of
\begin{equation}
{{m_b}\over {m_{\tau}}}\ltap {{\eta ^{1/2}}\over {x}}{{\eta _b}\over
{\eta _{\tau}}}\;, \label{blimit}
\end{equation}

\noindent {\bf B. Two-loop numerical results}

When the two-loop RGEs are considered, analytic solutions must be abandoned,
but the same qualitative behavior is found in the numerical
solutions. Furthermore, there is now the possibility that
the bottom quark Yukawa coupling at the
GUT scale becomes non-perturbative for large values of $\tan \beta$.
In our analysis we solve the two-loop RGEs of Eqs.~(\ref{dgidt}-\ref{dytaudt})
numerically\cite{CERN}, retaining all Yukawa couplings from the third
generation.

First we choose a value of $\alpha _3(M_Z)$ that is consistent with
experimental determinations and the preceding one-loop or two-loop
evolution of the gauge couplings in the absence of Yukawa couplings.
Specifically we take $\alpha _3(M_Z)=0.11$ or $\alpha _3(M_Z)=0.12$,
to bracket the indicated $\alpha _3(M_Z)$ range. For each particular
$\alpha _3(M_Z)$ we consider a range of
values for $\tan \beta$ and $m_b(m_b)$. For each choice of $\alpha _3(M_Z)$,
$\tan \beta $, $m_b$ we choose an input value of $m_t$. The Yukawa couplings
at scale $m_t$ are then given by
\begin{equation}
\lambda _t(m_t)={{\sqrt{2}m_t(m_t)}\over {v\sin \beta}}\;, \quad
\lambda _b(m_t)={{\sqrt{2}m_b(m_b)}\over {\eta _bv\cos \beta}}\;, \quad
\lambda _{\tau}(m_t)={{\sqrt{2}m_{\tau}(m_{\tau})}
\over {\eta _{\tau}v\cos \beta}}\;,
\label{Yukawa}
\end{equation}
and the $\alpha _i(m_t)$ are determined by Eqs.~(7)
and
(\ref{alpha3}) from the central values in Eq.~(6)
We take\cite{PDB} $m_{\tau}=1.784$ GeV.
The running of the vacuum expectation value $v$ between the fermion mass
scales is negligible for the range
of fermion masses considered here\cite{Ramond3}.
Starting at the scale $m_t$, we integrate the RGEs to the GUT scale, defined
to be the scale at which $\alpha _1(\mu )$ and $\alpha _2(\mu )$ intersect.
We then check to see if the equality
$\lambda _b(M_G^{})=\lambda _{\tau}(M_G^{})$ holds to within 0.01\%.
If the $b$ and $\tau $ Yukawas satisfy this condition, the solution is
accepted. If not, we choose another value of $m_t$ and repeat the
integration. Since our primary motivation here is to study the influence of
the $\alpha _3(M_Z)$ value on the Yukawa couplings, we do not enforce the
requirement that $\alpha _3(M_G^{})$ is equal to $\alpha _1(M_G^{})$
and $\alpha _2(M_G^{})$. Nevertheless the equality of $\alpha _1$,
$\alpha _2$, and $\alpha _3$ at $M_G^{}$ is typically violated by
$\ltap $ 4\% (2\%) for $\alpha _3(M_Z)=0.11\ (0.12)$. Such discrepancies
could easily exist from threshold effects at the
GUT scale\cite{barhall,threshold}.

We also explore the effects of taking the SUSY scale above $m_t$. We proceed as
described above, integrating the following
two-loop standard model RGEs numerically
from the top mass to the SUSY scale:
\begin{equation}
{{dg_i}\over {dt}}={g_i\over{16\pi^2 }}\left [b_i^{SM}g_i^2+{1\over {16\pi^2 }}
\left (\sum _{j=1}^3b_{ij}^{SM}g_i^2g_j^2
-\sum _{j=t,b,{\tau}}a_{ij}^{SM}g_i^2\lambda _j^2\right )\right ] \;,
\end{equation}
\begin{eqnarray}
{{d\lambda _t}\over {dt}}={{\lambda _t}\over {16\pi ^2}}
\Bigg[\Big [&-&\sum c_i^{SM}g_i^2+{3\over 2}\lambda _t^2-{3\over 2}
\lambda _b^2
+Y_2(S)\Big ] \nonumber \\
&+&{1\over {16\pi ^2}}\Bigg ({1187\over 600}g_1^4-{23\over 4}g_2^4-108g_3^4
-{9\over 20}g_1^2g_2^2+{19\over 15}g_1^2g_3^2+9g_2^2g_3^2\nonumber \\
&&\;\;\;\;\;\;\;+\left ({223\over 80}g_1^2+{135\over 16}g_2^2+16g_3^2\right )
\lambda _t^2-\left ({43\over 80}g_1^2-{9\over 16}g_2^2+16g_3^2
\right )\lambda _b^2\nonumber \\
&&\;\;\;\;\;\;\;+{5\over 2}Y_4(S)
-2\lambda \left (3\lambda _t^2+\lambda _b^2\right )\nonumber \\
&&\;\;\;\;\;\;\;+{3\over 2}\lambda _t^4
-{5\over 4}\lambda _t^2\lambda _b^2
+{11\over 4}\lambda _b^4\nonumber \\
&&\;\;\;\;\;\;\;+Y_2(S)\left ({5\over 4}\lambda _b^2
-{9\over 4}\lambda _t^2\right )-\chi _4(S)+{3\over 2}\lambda ^2
\Bigg )\Bigg ] \nonumber\;, \\
\label{dytsmdt}
\end{eqnarray}
\begin{eqnarray}
{{d\lambda _b}\over {dt}}={{\lambda _b}\over {16\pi ^2}}
\Bigg [\Big [&-&\sum c_i^{\prime SM}g_i^2+{3\over 2}\lambda _b^2
-{3\over 2}\lambda _t^2
+Y_2(S)\Big ] \nonumber \\
&+&{1\over {16\pi ^2}}\Bigg (-{127\over 600}g_1^4-{23\over 4}g_2^4-108g_3^4
-{27\over 20}g_1^2g_2^2+{31\over 15}g_1^2g_3^2+9g_2^2g_3^2\nonumber \\
&&\;\;\;\;\;\;\;-\left ({79\over 80}g_1^2-{9\over 16}g_2^2+16g_3^2\right )
\lambda _t^2+\left ({187\over 80}g_1^2+{135\over 16}g_2^2+16g_3^2
\right )\lambda _b^2\nonumber \\
&&\;\;\;\;\;\;\;+{5\over 2}Y_4(S)
-2\lambda \left (\lambda _t^2+3\lambda _b^2\right )\nonumber \\
&&\;\;\;\;\;\;\;+{3\over 2}\lambda _b^4
-{5\over 4}\lambda _b^2\lambda _t^2
+{11\over 4}\lambda _t^4\nonumber \\
&&\;\;\;\;\;\;\;+Y_2(S)\left ({5\over 4}\lambda _t^2
-{9\over 4}\lambda _b^2\right )-\chi _4(S)+{3\over 2}\lambda ^2
\Bigg )\Bigg ] \nonumber\;, \\
\label{dybsmdt}
\end{eqnarray}
\begin{eqnarray}
{{d\lambda _{\tau}}\over {dt}}={{\lambda _{\tau}}\over {16\pi ^2}}
\Bigg [\Big [&-&\sum c_i^{\prime \prime SM}g_i^2+{3\over 2}\lambda _{\tau}^2
+Y_2(S)\Big ] \nonumber \\
&+&{1\over {16\pi ^2}}\Bigg ({1371\over 200}g_1^4-{23\over 4}g_2^4
+{27\over 20}g_1^2g_2^2\nonumber \\
&&\;\;\;\;\;\;\;+\left ({387\over 80}g_1^2+{135\over 16}g_2^2\right )
\lambda _{\tau}^2+{5\over 2}Y_4(S)-6\lambda \lambda _{\tau}^2
\nonumber \\
&&\;\;\;\;\;\;\;+{3\over 2}\lambda _{\tau}^4
-{9\over 4}Y_2(S)\lambda _{\tau}^2-\chi _4(S)+{3\over 2}\lambda ^2
\Bigg )\Bigg ]\;,
\label{dyesmdt}
\end{eqnarray}
\begin{eqnarray}
{{d\lambda}\over {dt}}={1\over {16\pi ^2}}
\Bigg [&\Bigg \{&{9\over 4}\left ({3\over 25}g_1^4+{2\over 5}g_1^2g_2^2
+g_2^4\right )-\left ({9\over 5}g_1^2+9g_2^2\right )\lambda
+4Y_2(S)\lambda -4H(S)+12\lambda ^2\Bigg \} \nonumber \\
&+&{1\over {16\pi ^2}}\Bigg (-78\lambda ^3+18\left ({3\over 5}g_1^2+3g_2^2
\right )\lambda ^2+\left (-{73\over 8}g_2^4+{117\over 20}g_1^2g_2^2
+{1887\over 200}g_1^4\right )\lambda \nonumber \\
&&\;\;\;\;\;\;\;+{305\over 8}g_2^6-{867\over 120}g_1^2g_2^4
-{1677\over 200}g_1^4g_2^2-{3411\over 1000}g_1^6 \nonumber \\
&&\;\;\;\;\;\;\;-64g_3^2(\lambda _t^4
+\lambda _b^4)\nonumber \\
&&\;\;\;\;\;\;\;-{8\over 5}g_1^2(2\lambda _t^4
-\lambda _b^4+3\lambda _{\tau}^4]-{3\over 2}g_2^4Y_2(S)
+10\lambda Y_4(S)
\nonumber \\
&&\;\;\;\;\;\;\;+{3\over 5}g_1^2\left [
\left (-{57\over 10}g_1^2+21g_2^2\right )\lambda _t^2
+\left ({3\over 2}g_1^2+9g_2^2\right )\lambda _b^2\right .
\nonumber \\
&&\;\;\;\;\;\;\;\;\;\;\;\;\;\;\;\;\;\;
+\left .\left (-{15\over 2}g_1^2+11g_2^2\right )\lambda _{\tau}^2
\right ]\nonumber \\
&&\;\;\;\;\;\;\;-24\lambda ^2Y_2(S)-\lambda H(S)+6\lambda
\lambda _t^2\lambda_b^2\nonumber \\
&&\;\;\;\;\;\;\;+20\left [3\lambda _t^6
+3\lambda _b^6+\lambda _{\tau}^6\right ]\nonumber \\
&&\;\;\;\;\;\;\;-12\left [\lambda _t^4\lambda _b^2+\lambda _t^2\lambda _b^4
\right ]
\Bigg )\Bigg ]\;.
\label{dlmsmdt}
\end{eqnarray}
Here
\begin{equation}
Y_2(S)=3\lambda _t^2+3\lambda _b^2+\lambda _{\tau}^2\;,
\end{equation}
\begin{equation}
Y_4(S)={1\over 3}\left [3\sum c_i^{SM}g_i^2\lambda _t^2+3\sum c_i^{\prime SM}
g_i^2\lambda _b^2+\sum c_i^{\prime \prime SM}
g_i^2\lambda _{\tau}^2\right ]\;,
\end{equation}
\begin{equation}
\chi _4(S)={9\over 4}\left [3\lambda _t^4+
3\lambda _b^4+\lambda _{\tau}^4-{2\over 3}\lambda _t^2\lambda _b^2\right ]\;,
\end{equation}
\begin{equation}
H(S)=3\lambda _t^4+3\lambda _b^4+\lambda _{\tau}^4\;,
\end{equation}
and the coefficients $a^{SM}_{}$, $b^{SM}_{}$ and $c^{SM}_{}$ are
given in the appendix along with the full
matrix structure.

The initial values for $\alpha _3(M_Z)$,
$m_b$ and $m_t$ are chosen as before; in addition we are required to specify
the initial value of the quartic Higgs coupling $\lambda $ at scale $m_t$.
The Yukawa couplings at scale $m_t$ are
\begin{equation}
\lambda _t(m_t)={{\sqrt{2}m_t(m_t)}\over {v}}\;, \quad
\lambda _b(m_t)={{\sqrt{2}m_b(m_b)}\over {\eta _bv}}\;, \quad
\lambda _t(m_t)={{\sqrt{2}m_{\tau}(m_{\tau})}\over {\eta _{\tau}v}}\;,
\label{Yukawa_sm}
\end{equation}
and the $\alpha _i(m_t)$ are given by Eqs.~(6)-(8).
After integrating to the SUSY scale
we require that the matching condition
\begin{equation}
\lambda (M_{SUSY}^-)={1\over 4}\left ({3\over 5}g_1^2(M_{SUSY}^+)
+g_2^2(M_{SUSY}^+)\right )
\cos ^22\beta\;, \label{matching}
\end{equation}
is satisfied
to within 0.1\%. This condition\cite{Ramond1,HH}
results from integrating out the heavy Higgs
doublet at $M_{SUSY}^{}$. Below this scale only a Standard Model Higgs remains
with its quartic coupling given by Eq.~(\ref{matching}).
We also apply the matching conditions
\begin{eqnarray}
g_i(M_{SUSY}^-)&=&g_i(M_{SUSY}^+)\;, \label{m1}\\
\lambda _t(M_{SUSY}^-)&=&\lambda _t(M_{SUSY}^+)\sin \beta\;, \\
\lambda _b(M_{SUSY}^-)&=&\lambda _b(M_{SUSY}^+)\cos \beta\;, \\
\lambda _{\tau}(M_{SUSY}^-)&=&\lambda _{\tau}(M_{SUSY}^+)\cos \beta\;.
\label{m2}
\end{eqnarray}
If Eq.~(\ref{matching}) is not satisfied we choose another input
value of $\lambda (m_t)$ and repeat the process. We allow
$\tan \beta$ to span a wide grid of values.
After obtaining a satisfactory value of $\lambda $ that meets the boundary
condition above, we integrate the two-loop SUSY RGEs to the GUT scale,
defined by the equality
$\alpha _1(M_G)=\alpha _2(M_G)$. At the GUT scale we require
$\lambda _b(M_G)=\lambda _{\tau}(M_G)$ to within 0.1\%. If this condition is
not met, we repeat the entire process, choosing other initial
values for $m_t$ and
$\lambda $.

The parameter $\beta $ also runs in going
from the SUSY scale to the electroweak
scale\cite{HH}. However this effect is small and we neglect it here.

In Figure 5 the resulting contours of constant $m_b$ are given in the
$m_t,\tan \beta$ plane\cite{Ramond1,KLN} for the choices of
$\alpha _3(M_Z)=0.11$ and 0.12 and the supersymmetry scales $M_{SUSY}^{}=m_t$
and 1 TeV.
The contours shown are
$m_b=4.1, 4.25, 4.4$ GeV (corresponding to the central value
of $m_b$ and its 90\%
confidence range from the GL QCD sum rule
determination) and $m_b=5.0$ GeV (representing a typical
constituent $b$-quark mass value). For a given $m_b$ and $m_t\ltap 175$ GeV,
there is a high solution and a low solution for $\tan \beta$
as anticipated in \S IIIa. Thus, once $m_t$
is experimentally known and the choice of $m_b$ resolved by other
considerations (such as the CKM matrix elements addressed
subsequently), the assumption of Yukawa unification at the GUT scale will
select two possible values for $\tan \beta$. For example for $m_t=150$ GeV
and $m_b=4.25$ GeV, the solutions with $\alpha _3(M_Z)=0.11$ are
\begin{equation}
\tan \beta =1.35\quad {\rm or} \quad \tan \beta =56\;.
\end{equation}
For $m_t\ltap 175$ GeV the low solution is well-approximated by
\begin{equation}
\sin \beta =0.78\left ({{m_t}\over {150 \;{\rm GeV}}}\right )\;.
\end{equation}
Such knowledge of $\tan\beta$ would greatly simplify SUSY Higgs
analyses\cite{BBSP}. Without imposing any other constraints, the top
quark mass $m_t$ can be arbitrarily small.

The plots rise very steeply for the maximal value of $m_t$. This results
because the linear relation exhibited in Eq.~(\ref{linear}) and in the plot
in Ref.~\cite{BBHZ} between $m_t$ and
$\sin \beta$ is mapped into a vertical line for sufficiently large $\tan \beta$
($\gtap 10$). The deviation of these contours from being strictly vertical
results from the contributions of $\lambda _b$ and $\lambda _{\tau}$ to the
Yukawa coupling evolution.

An upper limit on $m_t$ is determined entirely by the $m_b/m_{\tau}$ ratio.
We find the $m_t$ upper limits shown in Table 3 for the two choices of
$\alpha _3(M_Z)$. It is interesting that the predicted upper limit for $m_t$
coincides with that allowed by electroweak radiative corrections\cite{PDB}.
\vskip 0.4in
{\center \begin{tabular}{|c|c|c|}
\hline
\multicolumn{1}{|c|}{}
&\multicolumn{2}{|c|}{$\alpha _3(M_Z)$}
\\ \cline{2-3}
\multicolumn{1}{|c|}{$M_{SUSY}^{}$}
&\multicolumn{1}{|c|}{0.11}
&\multicolumn{1}{|c|}{0.12}
\\ \hline \hline
\multicolumn{1}{|c|}{$m_t$}
&\multicolumn{1}{|c|}{187}
&\multicolumn{1}{|c|}{193}
\\ \hline
\multicolumn{1}{|c|}{1 TeV}
&\multicolumn{1}{|c|}{192}
&\multicolumn{1}{|c|}{199}
\\ \hline
\end{tabular}
\vskip .3in }
\begin{center}
{\bf Table 3:}  Maximum values of $m_t(m_t)$ in GeV consistent with\\
the 90\% confidence levels of the $m_b(m_b)$ values of GL.
\end{center}
\vskip .4in

Our contours of $m_b/m_{\tau}$ in Fig. 5 have about a 10\% higher $m_b$
than those given in
Ref.~\cite{KLN} presumably because they employed the one-loop QCD results for
the scaling factors $\eta _f$ with the two-loop expression for $\alpha _3$
rather than the three-loop QCD for both $\eta _f$ and $\alpha _3$ that we use
here.

As $\alpha _3(M_Z)$ gets larger, smaller values of $y$ are needed
to obtain obtain the correct $m_b/m_{\tau}$ ratio. In turn
larger values of $\lambda _t(\mu )$ are needed via Eq.~(\ref{defy}).
For $\alpha _3(M_Z)\gtap 0.12$
and $m_b\ltap 4.2$ GeV, the value of $\lambda _t(\mu )$ near the
GUT scale can be driven into the nonperturbative regime. In Figure 6 we show
the values of $\lambda _t(M_G)$ and $\lambda _b(M_G)$ obtained for the
solutions in Fig. 5. Fixed points in the quark Yukawa couplings exist at
$\lambda \approx 1$, so a Yukawa coupling only slightly larger than the
fixed point at the
scale $m_t$ can diverge as it is evolved to the GUT scale.
For large values of the Yukawa couplings the two-loop contributions
to the RGEs contribute
a fraction $x$ of the one-loop contributions when
\begin{eqnarray}
\lambda _t&=&\sqrt{{{6(16\pi ^2x)}}\over {22}}\approx 6.5\sqrt{x}
\;, \label{tlim} \\
\lambda _b&=&\sqrt{{{7(16\pi ^2x)}}\over {28}}\approx 6.3\sqrt{x}
\;,\label{blim}
\end{eqnarray}
as can be deduced from Eqs.~(\ref{dytdt}) and (\ref{dybdt}).
When $x\approx 1$ we are clearly in the nonperturbative regime.
If we adopt the criteria that the two-loop effects always be less
than a quarter of the one-loop effects, then $\lambda _t$ and $\lambda _b$
are nonperturbative when they remain below 3.3 and 3.1 respectively all the way
to the GUT scale. This is true for all of the curves presented in Figure 6,
except for the $m_b=4.1$ GeV contours for $\alpha _3(M_Z)=0.12$;
hence the exact position of this
contour cannot be
predicted with accuracy.

In Figure 7 we show the evolution of the Yukawa couplings from the SUSY scale
to the GUT scale. The nonperturbative regime for the case discussed above
occurs only near the GUT scale.

In some SO(10) GUT models the top quark Yukawa coupling $\lambda _t$ is
unified with the $\lambda _b$ and $\lambda _{\tau}$ at the GUT scale.
Imposing this constraint selects a unique value for $\tan \beta$ and $m_t$.
This solution is given by the intersection of $\lambda _t(M_G)$ and
$\lambda _b(M_G)$, which occurs for large
$\tan \beta\gtap 50$: see Fig. 6.

One could also consider the unification of the Yukawa couplings at some scale
other than that at which the gauge couplings unify\cite{Ramond1,KLN}. Since
$R_{b/\tau}$
increases as it evolves from the GUT scale to the electroweak scale, Yukawa
unification at a scale larger than the gauge coupling unification scale gives
a larger $m_b/m_{\tau}$ ratio.

The authors of Ref.~\cite{Ramond1} predict the light physical Higgs mass rather
precisely. However this prediction is related to their assumption
(and the one we use here)
that the heavy Higgs
doublet is integrated out at $M_{SUSY}^{}$. This means that the
heavy physical Higgs bosons have masses $M_H\approx M_A\approx M_{H^{\pm}}
\approx M_{SUSY}^{}>>M_Z$, which requires that
the light Higgs mass is close to
its upper limit. The relation of $\sin \beta$ to $m_t$ then
fixes the one-loop corrections to the light Higgs mass.

\section{Fermion Mass Ansatz}

By assuming an ansatz for Yukawa matrices at the GUT scale and evolving these
matrices down to the electroweak scale, predictions can be obtained for the
quark and lepton masses and the CKM matrix
elements\cite{Ramond1,DHR,Giudice,Ramond2}. Much work has been done on
individual relations such as $|V_{ud}|\approx \sqrt{m_s/m_d}$ and
$|V_{cb}|\approx \sqrt{m_c/m_t}$ which are imposed
at the GUT scale as described in \S III.
Recently interest has been revived in models that
involve several such relations, leading to
a number of predictions for quark masses and CKM matrix elements at the
electroweak scale\cite{Ramond1,DHR,Giudice,genmat}.
 The relations evolve according to RGEs, and the main
effects are determined by the largest couplings. For moderate values of
$\tan \beta$ (i.e. $\tan \beta \ltap 10$),
these are the gauge couplings $g_i$ and the top quark Yukawa
coupling $\lambda _t$. For large values of $\tan \beta (\approx m_t/m_b)$ the
effects of $\lambda _b$ and $\lambda _{\tau}$ can also be significant.
Various individual relations at the GUT scale
such as $|V_{cb}|\approx \sqrt{m_c/m_t}$
can be satisfied for certain choices of these
Yukawa couplings. The remarkable aspect of these fermion mass ans\"{a}tze is
that many relations can be
made to work at one time. We shall concentrate in this section on two
predictive ways of generating mixing between the second and third
generations which put those mixing contributions entirely
in the up quark Yukawa matrix\cite{Ramond1,DHR,Ramond2} or entirely in the
down quark Yukawa matrix\cite{Giudice}.

\newpage
\noindent {\bf A. The HRR/DHR Model}

Harvey, Ramond and Reiss\cite{Ramond2}
proposed that the Yukawa matrices at the
GUT scale have the form
\begin{equation}
{\bf U}= \left( \begin{array}{c@{\quad}c@{\quad}c}
0 & C & 0 \\ C & 0 & B \\ 0 & B & A
\end{array} \right) \qquad
{\bf D} = \left( \begin{array}{c@{\quad}c@{\quad}c}
0 & Fe^{i\phi} & 0 \\ Fe^{-i\phi} & E & 0 \\ 0 & 0 & D
\end{array} \right) \nonumber \;, \label{HRR/DHRyuk1}
\end{equation}
\begin{equation}
{\bf E} = \left( \begin{array}{c@{\quad}c@{\quad}c}
0 & F & 0 \\ F & -3E & 0 \\ 0 & 0 & D
\end{array} \right) \;. \qquad
\label{HRR/DHRyuk2}
\end{equation}
These matrices incorporate both Fritzsch zeros\cite{Fritzsch} and
the Georgi-Jarlskog relation\cite{GJ} between down quark and charged
lepton matrix elements.
This relative factor of three has been realized in Higgs models with certain
vacuum breaking patterns. HRR obtained the above ansatz using a
{\bf 10} and three {\bf 126} Higgs multiplets in
an SO(10) GUT model to obtain various relationships between CKM matrix elements
and quark masses.
The GUT ansatz of Eqs.~(\ref{HRR/DHRyuk1}) and (\ref{HRR/DHRyuk2})
is also the basis for the recent RGE analysis by Dimopoulos,
Hall and  Raby\cite{DHR}. Henceforth we shall refer to this ansatz as the
HRR/DHR model. It yields the relation
$|V_{cb}|=\sqrt{\lambda _c/\lambda _t}$ at the GUT scale.

Renormalization group evolution generates non-zero entries in the above Yukawa
matrices and also splits $B_1\equiv {\bf U}_{23}$ and
$B_2\equiv {\bf U}_{32}$ to give the matrices at the electroweak scale
of the form
\begin{equation}
{\bf U}= \left( \begin{array}{c@{\quad}c@{\quad}c}
0 & C & 0 \\ C & \delta _u & B_1 \\ 0 & B_2 & A
\end{array} \right) \qquad
{\bf D} = \left( \begin{array}{c@{\quad}c@{\quad}c}
0 & Fe^{i\phi} & 0 \\ Fe^{-i\phi} & E & \delta _d \\ 0 & 0 & D
\end{array} \right) \nonumber \;,
\end{equation}
\begin{equation}
{\bf E} = \left( \begin{array}{c@{\quad}c@{\quad}c}
0 & F^{\prime} & 0 \\ F^{\prime} & -3E^{\prime} & 0 \\ 0 & 0 & D^{\prime}
\end{array} \right) \;. \qquad
\label{HRR/DHRyuk}
\end{equation}
%
The quantities $A$, $D$ and $D^{\prime}$ are equivalent to $\lambda _t$,
$\lambda _b$ and $\lambda _{\tau}$ respectively up to subleading corrections
in the mass matrix diagonalization.
The one-loop solutions\cite{DHR} to leading
order in the hierarchy can be obtained analytically neglecting
$\lambda _b$ and $\lambda _{\tau}$. The one-loop results for the
CKM elements at the scale $m_t$ are
\begin{eqnarray}
|V_{us}|&=&\left [{{\eta _sm_d}\over {\eta _dm_s}}
+{{\eta _cm_u}\over {\eta _um_c}}+2\sqrt{{{\eta _s\eta_cm_um_d}\over
{\eta_d\eta _um_sm_c}}}\cos \phi\right ]^{1/2}\;, \label{Vusdhr} \\
|V_{cb}|&=&\sqrt{{{ym_c}\over {\eta _cm_t}}}\;, \\
\left |{{V_{ub}}\over {V_{cb}}}\right |&=&\sqrt{{{\eta _cm_u}
\over {\eta _um_c}}}\;,
\end{eqnarray}
where $\eta _i(m_t)$ is defined by Eq.~(\ref{etadef})
and $y(m_t)$ by Eq.~(\ref{defy}).
The angle $\phi $ is \'a priori arbitrary.
The down-type quark masses are related to the corresponding lepton masses by
\begin{eqnarray}
m_d&=&{{\eta ^{1/2}}\over {x}}{{\eta _d}\over {\eta _e}}
3m_e \;, \\
m_s&=&{{\eta ^{1/2}}\over {x}}{{\eta _s}\over {\eta _{\mu}}}
{{m_{\mu}}\over {3}} \;, \\
m_b&=&y{{\eta ^{1/2}}\over {x}}{{\eta _b}\over {\eta _{\tau}}}m_{\tau}\;.
\label{mbmtaudhr}
\end{eqnarray}

Using the general expressions for the two-loop RGEs given in the appendix and
keeping only terms unsuppressed by the hierarchy, one obtains
Eqs.~(\ref{dgidt})--(\ref{dytaudt}) as well as
\begin{eqnarray}
{{dB_1}\over {dt}}={{B_1}\over {16\pi ^2}}
\Bigg [\Bigg (&-&\sum c_ig_i^2+6\lambda _t^2+
{{\lambda _t\lambda _b\delta _d}\over {B_1}}
\Bigg )
\nonumber \\
&+&{1\over {16\pi ^2}}\Bigg (\sum \left(c_ib_i+c_i^2/2\right )g_i^4
+g_1^2g_2^2+{136\over 45}g_1^2g_3^2+8g_2^2g_3^2\nonumber \\
&&\;\;\;\;\;\;+\lambda _t^2\left({6\over 5}g_1^2+6g_2^2
+16g_3^2\right )
+{2\over 5}{{\lambda _t\lambda _b\delta _d}\over {B_1}}g_1^2
\nonumber \\
&&\;\;\;\;\;\;-\left \{22\lambda _t^4+5\lambda _t^2\lambda _b^2
+{{\lambda _t\lambda _b\delta _d}\over {B_1}}\left (5\lambda _b^2
+\lambda _{\tau}^2\right )\right \}\Bigg )\Bigg ] \;, \label{dB1dt}
\end{eqnarray}
\begin{eqnarray}
{{dB_2}\over {dt}}={{B_2}\over {16\pi ^2}}
\Bigg [\Bigg (&-&\sum c_ig_i^2+6\lambda _t^2
+\lambda _b^2\Bigg )
\nonumber \\
&+&{1\over {16\pi ^2}}\Bigg (\sum \left(c_ib_i+c_i^2/2\right )g_i^4
+g_1^2g_2^2+{136\over 45}g_1^2g_3^2+8g_2^2g_3^2\nonumber \\
&&\;\;\;\;\;\;+\lambda _t^2\left({6\over 5}g_1^2+6g_2^2
+16g_3^2\right )+{2\over 5}\lambda _b^2g_1^2
\nonumber \\
&&\;\;\;\;\;\;-\left \{22\lambda _t^4+5\lambda _t^2\lambda _b^2
+5\lambda _b^4
+\lambda _b^2\lambda _{\tau}^2\right \}\Bigg )\Bigg ] \;, \label{dB2dt}
\end{eqnarray}
\begin{eqnarray}
{{d\delta _u}\over {dt}}={{\delta _u}\over {16\pi ^2}}
\Bigg [\Bigg (&-&\sum c_ig_i^2+3\lambda _t^2+3
{{\lambda _tB_1B_2}\over {\delta _u}}
+{{\lambda _b\delta _dB_2}\over {\delta _u}}
\Bigg )
\nonumber \\
&+&{1\over {16\pi ^2}}\Bigg (\sum \left(c_ib_i+c_i^2/2\right )g_i^4
+g_1^2g_2^2+{136\over 45}g_1^2g_3^2+8g_2^2g_3^2\nonumber \\
&&\;\;\;\;\;\;+\lambda _t^2\left({4\over 5}g_1^2+16g_3^2\right )
+{{\lambda _tB_1B_2}\over {\delta _u}}\left({2\over 5}g_1^2
+6g_2^2\right )
+{2\over 5}{{\lambda _b\delta _dB_2}\over {\delta _u}}g_1^2
\nonumber \\
&&\;\;\;\;\;\;-\left \{9\lambda _t^4+3\lambda _t^2\lambda _b^2
+{{\lambda _tB_1B_2}\over {\delta _u}}\left (13\lambda _t^2
+2\lambda _b^2\right )
+{{\lambda _b\delta _dB_2}\over {\delta _u}}\left (5\lambda _b^2
+\lambda _{\tau}^2\right )
\right \}\Bigg )\Bigg ] \nonumber \;, \label{ddeltaudt} \\
\end{eqnarray}
\begin{eqnarray}
{{d\delta _d}\over {dt}}={{\delta _d}\over {16\pi ^2}}
\Bigg [\Bigg (&-&\sum c_i^{\prime}g_i^2+6\lambda _b^2
+\lambda _{\tau}^2+
{{\lambda _t\lambda _bB_1}\over {\delta _d}}
\Bigg )
\nonumber \\
&+&{1\over {16\pi ^2}}\Bigg (\sum \left(c_i^{\prime}b_i+c_i^{\prime 2}
/2\right )g_i^4
+g_1^2g_2^2+{8\over 9}g_1^2g_3^2+8g_2^2g_3^2\nonumber \\
&&\;\;\;\;\;\;+\lambda _b^2\left({2\over 5}g_1^2+6g_2^2
+16g_3^2\right )
+{6\over 5}\lambda _{\tau}^2g_1^2
+{4\over 5}{{\lambda _t\lambda _bB_1}\over {\delta _d}}g_1^2
\nonumber \\
&&\;\;\;\;\;\;-\left \{22\lambda _b^4+5\lambda _t^2\lambda _b^2
+3\lambda _b^2\lambda _{\tau}^2+3\lambda _{\tau}^4
+{{\lambda _t\lambda _bB_1}\over {\delta _d}}5\lambda _t^2
\right \}\Bigg )\Bigg ] \;.\label{ddeltaddt}
\end{eqnarray}
Notice that since $1/B_2 dB_2/dt$= $1/\lambda _t d\lambda_t/dt$, the ratio
$B_2/\lambda _t$ is constant over all scales and is in particular equal to its
value at the GUT scale ($B_{2G}/\lambda _t(M_G^{})$).

With these RGEs we can include the additional experimental constraints from
the charm mass $m_c$ and the CKM matrix element $|V_{cb}|$
to determine the allowed
region of the HRR/DHR model in the $m_t,\tan \beta$ plane. An analysis at the
one-loop level neglecting $\lambda _b$ and $\lambda _{\tau}$ relative to
$\lambda _t$ was presented in Ref.~\cite{BBHZ}.

The Yukawa matrices are diagonalized by unitary matrices $V_u^L, V_u^R,
V_d^L, V_d^R$ so that ${\bf U^{diag}}=V_u^L{\bf U}V_u^{R\dagger}$ and
${\bf D^{diag}}=V_d^L{\bf D}V_d^{R\dagger}$. The CKM matrix is then given by
$V_{CKM}^{}=V_u^LV_d^{L\dagger }$. We define a ``running'' CKM matrix by
diagonalizing the Yukawa matrices {\bf U} and {\bf D} at any scale $t$. We
find that $\lambda _c/\lambda _t$ and $|V_{cb}|$ are described in terms of
the Yukawa matrices by
\begin{equation}
R_{c/t}\equiv {{\lambda _c}\over{\lambda _t}}
=\left ({{B_1B_2}\over{\lambda _t^2}}
-{{\delta _u}\over {\lambda _t}}\right ) \;,
\end{equation}
\begin{equation}
|V_{cb}|={{B_1}\over {\lambda _t}}-{{\delta _d}\over {\lambda _b}} \;,
\end{equation}
with
\begin{equation}
{{m_c}\over {m_t}}=\eta _cR_{c/t}(m_t)\label{mcmt}\;.
\end{equation}
To leading order in the mass hierarchy, the ratio $R_{c/t}$ is given by
the ratio of eigenvalues of the $2\times 2$ submatrix of ${\bf U}$
in the second and third
generations while $V_{cb}$ is given by the difference in the rotation
angles needed to rotate away the upper right hand entry in the submatrices of
${\bf U}$ and ${\bf D}$.
Given that the mass hierarchies exist, there is a simple iterative numerical
procedure for diagonalizing the mass matrices ${\bf U}$ and ${\bf D}$ and
obtaining the CKM matrix. We have checked that the corrections to the above
formulas from contributions subleading in the mass hierarchy are small.

It is straightforward to derive the resulting renormalization group
equations from Eqs.~(\ref{dB1dt})-(\ref{ddeltaddt})
\begin{eqnarray}
{{dR_{c/t}}\over {dt}}=-{{R_{c/t}}\over {16\pi ^2}}
&\Bigg [&\left (3\lambda _t^2
+\lambda _b^2\right )
\nonumber \\
&+&{1\over {16\pi ^2}}\left (
\lambda _t^2\left ({2\over 5}g_1^2+6g_2^2\right )
+{2\over 5}\lambda _b^2g_1^2
-\left (13\lambda _t^4+2\lambda _t^2\lambda _b^2
+5\lambda _b^4
+\lambda _b^2\lambda _{\tau}^2\right )\right )\Bigg] \;, \label{dRppdt} \\
{{d|V_{cb}|}\over {dt}}=-{{|V_{cb}|}\over {16\pi ^2}}
&\Bigg [&\left (\lambda _t^2
+\lambda _b^2\right )
\nonumber \\
&+&{1\over {16\pi ^2}}\left (
{4\over 5}\lambda _t^2g_1^2
+{2\over 5}\lambda _b^2g_1^2
-\left (5\lambda _t^4
+5\lambda _b^4
+\lambda _b^2\lambda _{\tau}^2\right )\right )\Bigg ] \;, \label{dVcbdt}
\end{eqnarray}

The corresponding evolution equations in the
Standard Model are given by
\begin{eqnarray}
{{dR_{c/t}}\over {dt}}=-{{R_{c/t}}\over {16\pi ^2}}
&\Bigg [&\left ({3\over 2}\lambda _t^2
-{3\over 2}\lambda _b^2\right )
\nonumber \\
&+&{1\over {16\pi ^2}}\Bigg (
\lambda _t^2\left ({223\over 80}g_1^2+{135\over 16}g_2^2+16g_3^2\right )
-\lambda _b^2\left ({43\over 80}g_1^2-{9\over 16}g_2^2+16g_3^2\right )
\nonumber \\
&&\;\;\;\;\;\;\;\;\;\;-2\lambda (3\lambda _t^2+\lambda _b^2) \nonumber \\
&&\;\;\;\;\;\;\;\;\;\;-\left ({21\over 4}\lambda _t^4
+{17\over 4}\lambda _t^2\lambda _b^2
-{13\over 2}\lambda _b^4+{9\over 4}\lambda _t^2\lambda _{\tau}^2
-{5\over 4}\lambda _b^2\lambda _{\tau}^2\right )\Bigg )\Bigg] \;,
\label{dRppSMdt} \\
{{d|V_{cb}|}\over {dt}}={{|V_{cb}|}\over {16\pi ^2}}
&\Bigg [&\left ({3\over 2}\lambda _t^2
+{3\over 2}\lambda _b^2\right )
\nonumber \\
&+&{1\over {16\pi ^2}}\Bigg (
\lambda _t^2\left ({79\over 80}g_1^2-{9\over 16}g_2^2+16g_3^2\right )
+\lambda _b^2\left ({43\over 80}g_1^2-{9\over 16}g_2^2+16g_3^2\right )
\nonumber \\
&&\;\;\;\;\;\;\;\;\;\;+2\lambda (\lambda _t^2+\lambda _b^2) \nonumber \\
&&\;\;\;\;\;\;\;\;\;\;-\left ({13\over 2}\lambda _t^4+{11\over 2}
\lambda _t^2\lambda _b^2
+{13\over 2}\lambda _b^4+{5\over 4}\lambda _t^2\lambda _{\tau}^2
+{5\over 4}\lambda _b^2\lambda _{\tau}^2\right )\Bigg )\Bigg]
\;, \label{dVcbSMdt}
\end{eqnarray}
The evolution equations in Eqs.~(\ref{dRppSMdt})-(\ref{dVcbSMdt})
are obtained from the two-loop RGEs of the standard model
given in Ref~\cite{smrge} and in the appendix.

In the supersymmetric model $|V_{cb}|$ increases with the running
from the GUT scale to the
electroweak scale\cite{KSBabu}; this is evident at the two-loop level
in Eq.(\ref{dVcbdt}). The opposite behavior occurs in
Eq.~(\ref{dVcbSMdt}) for the nonsupersymmetric
Standard Model where $|V_{cb}|$ decreases
as the running mass decreases\cite{Ma}.
Fig. 8 shows the running of
$|V_{cb}|$ for the cases
$M_{SUSY}^{}=m_t$ and 1 TeV. In contrast to $|V_{cb}|$
the ratio $R_{c/t}$ increases monotonically
as the running mass decreases in both the Standard Model and
supersymmetric model cases.

We stress that Eqs.~(\ref{dRppdt}) and (\ref{dVcbSMdt})
are the correct evolution equations regardless of the fermion mass ansatz at
the GUT scale. Changing the ansatz just changes the boundary conditions at
the GUT scale (terms subleading in the mass hierarchy differ between
models, but this is a negligible effect).
In a model for which the relationship
$|V_{cb}|=\sqrt{\lambda _c/\lambda _t}$ holds
(as in the HRR/DHR model), this boundary condition is
$\sqrt{R_{c/t}(M_G)}=|V_{cb}(M_G)|$. In Giudice's model, to be described below,
the mixing between the second and third generations arises in the down quark
Yukawa matrix alone,
and so in his model $R_{c/t}$ and $|V_{cb}|$ are unrelated at the GUT scale.

In our analysis of the CKM constraints we proceed
as in the discussion of the
calculation for Figure 5. We numerically solve the two-loop RGEs as given by
Eqs.~(\ref{dgidt})-(\ref{dytaudt}),(\ref{dRppdt})-(\ref{dVcbdt}) for the case
$M_{SUSY}^{}=m_t$. As before,
we consider the representative choices $\alpha _3(M_Z)=0.11$ and
$\alpha _3(M_Z)=0.12$. For each $\alpha _3(M_Z)$ choice, we
consider a grid of $\tan \beta$ values, holding $|V_{cb}(m_t)|$
and $m_c$ fixed. We then choose input values for $m_t$
and $m_b$ (given $\alpha _3(M_Z)$, $\tan \beta$,
$|V_{cb}|$, $m_c$) in terms of which
all running parameters are uniquely specified at
$m_t$: $\lambda _t(m_t)$, $\lambda _b(m_t)$ and $\lambda _{\tau}(m_t)$ are
given by Eq.~(\ref{Yukawa}), $\alpha _i(m_t)$ are determined by
Eqs.~(7)
and (\ref{alpha3}) using the central values in Eq.~(6)
$R_{c/t}$ is given by Eq.~(\ref{mcmt}), and $|V_{cb}|$ at scale $m_t$ is an
input. After integrating the RGEs from $m_t$ to $M_G$ we check the
constraints
\begin{mathletters}
\begin{eqnarray}
\lambda _b(M_G)&=&\lambda _{\tau}(M_G)\;, \\
\sqrt{R_{c/t}(M_G)}&=&|V_{cb}(M_G)|\;.
\end{eqnarray}
\end{mathletters}
If either of these conditions is not satisfied to within 0.2\%,
we choose another input value for $m_t$ and $m_b$ and repeat the integration.

We also carry out the RGE calculations with a SUSY scale at 1 TeV. This is done
exactly as described in the previous section. In addition to the other
parameters, we choose an input value for the quartic Higgs coupling
$\lambda $ at scale $m_t$.
We then integrate the two-loop standard model RGEs to the
SUSY scale and require that Eq.~(\ref{matching}) hold to within 0.1\%. For
such solutions we apply the other appropriate boundary conditions
(given by Eqs.~(\ref{m1})-(\ref{m2})) and integrate the two-loop SUSY RGEs to
the GUT scale, where we require that $\lambda _b(M_G)=\lambda _{\tau}(M_G)$
and $\sqrt{R_{c/t}(M_G)}=|V_{cb}(M_G)|$ to within 0.2\%.
In our calculation
we require that $m_b, m_c$ and $|V_{cb}|$ be within the experimentally
determined 90\% confidence levels of the quark mass determinations of GL
($4.1 < m_b < 4.4$ GeV, $1.19 < m_c < 1.35$ GeV)
and the recent Particle Data Book value\cite{PDB}
for $|V_{cb}|$ ($0.032 < |V_{cb}| < 0.054$).

In Fig. 9 the contours of constant $|V_{cb}|$ are shown in the
$m_t,\tan \beta $ plane for a fixed $m_c=1.27$ GeV.
In Figs. 10 and 11  we show the contours obtained by applying
only the constraint in Eq.~(75a) as in Fig. 5 along with
the contours obtained by applying both Eqs.~(75a) and (75b)
for fixed $m_c$ as in Fig. 9.
In Fig. 10 the value of $m_c$ is fixed at 1.27 GeV
and contours of $|V_{cb}|$ are shown.
In Fig. 11 $|V_{cb}|$ is fixed at its maximum allowed experimental value of
$|V_{cb}|=0.054$ (at 90\% C.L.) and three values of
$m_c$ are plotted (corresponding to the central $m_c$ value and the 90\%
C.L. values from GL).

For large $\tan \beta$ the effects
of including $\lambda _b$ and $\lambda _{\tau}$ in the RGEs increase
$|V_{cb}|$. In order to satisfy $|V_{cb}| < 0.054$, the maximum allowed
value of $\tan \beta$
for $\alpha _3(M_Z)=0.11$ is about 50(60) for $M_{SUSY}=m_t (1 {\rm TeV})$;
see Fig. 11.
For this value of $\alpha _3(M_Z)$ the HRR/DHR model
predicts that $|V_{cb}|$ still
lies at the upper end of its allowed 90\% confidence
level range when the effects of $\lambda _b$ and
$\lambda _{\tau}$ at large $\tan \beta$ are included in the two-loop RGEs;
see Fig. 10.
Allowing $m_b$ to become
larger than the narrow window $m_b=4.1-4.4$ GeV requires bigger
$|V_{cb}|$ which is unacceptable.
The higher $b$ mass contour $m_b=5$ GeV is not consistent
with the GUT scale ansatz for $\alpha _3(M_Z)=0.11$.
The largest consistent values of $m_b$ are given in Table 4.
\vskip 0.2in
{\center \begin{tabular}{|c|c|c|}
\hline
\multicolumn{1}{|c|}{}
&\multicolumn{2}{|c|}{$\alpha _3(M_Z)$}
\\ \cline{2-3}
\multicolumn{1}{|c|}{$M_{SUSY}^{}$}
&\multicolumn{1}{|c|}{0.11}
&\multicolumn{1}{|c|}{0.12}
\\ \hline \hline
\multicolumn{1}{|c|}{$m_t$}
&\multicolumn{1}{|c|}{4.56}
&\multicolumn{1}{|c|}{5.28}
\\ \hline
\multicolumn{1}{|c|}{1 TeV}
&\multicolumn{1}{|c|}{4.70}
&\multicolumn{1}{|c|}{5.33}
\\ \hline
\end{tabular}
\vskip .15in }
\begin{center}
{\bf Table 4:}  Maximum values of $m_b(m_b)$ in GeV consistent with\\
the 90\% confidence levels of $|V_{cb}|$ and $m_c(m_c)$.
\end{center}
\vskip .6in

With $\alpha _3(M_Z)=0.12$,
$|V_{cb}|$ can be much closer to its central value,
enhancing the plausibility of the HRR/DHR model,
with the only caveat being that low
$m_b$ ($\ltap 4.2$ GeV)
values produce $\lambda _t(M_G)$
values which are close to being non-perturbative for most values of
$\tan \beta$: see Figs. 6b, 6d.
Notice that the dominant effect of taking the larger value of $\alpha _3(M_Z)$
indicated by two-loop evolution
is to increase the QCD-QED scaling factor
$\eta _c$, thereby allowing $|V_{cb}|$ to be
smaller and in better agreement with experiment.

Imposing the constraints on $m_b$, $m_c$ and $|V_{cb}|$
also gives the lower limits on the top quark mass
since the $|V_{cb}|$ contours in the smaller $\tan \beta $ region are
steeper and eventually cross the $m_b/m_{\tau}$ contours\cite{BBHZ}. These
lower limits on $m_t$ are summarized in Table 5.
\vskip 0.6in
{\center \begin{tabular}{|c|c|c|}
\hline
\multicolumn{1}{|c|}{}
&\multicolumn{2}{|c|}{$\alpha _3(M_Z)$}
\\ \cline{2-3}
\multicolumn{1}{|c|}{$M_{SUSY}^{}$}
&\multicolumn{1}{|c|}{0.11}
&\multicolumn{1}{|c|}{0.12}
\\ \hline \hline
\multicolumn{1}{|c|}{$m_t$}
&\multicolumn{1}{|c|}{155 (1.45)}
&\multicolumn{1}{|c|}{118 (0.75)}
\\ \hline
\multicolumn{1}{|c|}{1 TeV}
&\multicolumn{1}{|c|}{151 (1.16)}
&\multicolumn{1}{|c|}{116 (0.64)}
\\ \hline
\end{tabular}
\vskip .3in }
\begin{center}
{\bf Table 5:}  Minimum values of $m_t(m_t)$ ($\tan \beta$)
in GeV consistent with\\
the 90\% confidence levels of $m_b(m_b)$, $|V_{cb}|$ and $m_c(m_c)$.
\end{center}
\vskip .6in
The constraints on $m_b/m_{\tau}$, $|V_{cb}|$ and $m_c$
completely determine the
allowed region in the $m_t,\tan \beta$ plane of the HRR/DHR model.
Other constraints such as the
$\epsilon $ parameter for CP violation in the neutral kaon system,
$B$ mixing or the lighter quark masses affect only the
other parameters in the model\cite {BBHZ}.

If the Yukawa unification is assumed to occur at a scale higher than the gauge
couplings, then the predicted value for $|V_{cb}|$ will be
lower\cite{Ramond1}
and easier to reconcile with the experimental data.

\noindent {\bf B. The Giudice Model}

Giudice has proposed a different Yukawa mass ansatz\cite{Giudice}
of the form
\begin{equation}
{\bf U}= \left( \begin{array}{c@{\quad}c@{\quad}c}
0 & 0 & b \\ 0 & b & 0 \\ b & 0 & a
\end{array} \right) \qquad
{\bf D} = \left( \begin{array}{c@{\quad}c@{\quad}c}
0 & fe^{i\phi} & 0 \\ fe^{-i\phi} & d & nd \\ 0 & nd & c
\end{array} \right) \nonumber \;,
\end{equation}
\begin{equation}
{\bf E} = \left( \begin{array}{c@{\quad}c@{\quad}c}
0 & f & 0 \\ f & -3d & nd \\ 0 & nd & c
\end{array} \right) \qquad
\label{Guiyuk}
\end{equation}
This model uses a geometric mean relation
$m_c^2=m_um_t$ at the GUT scale
to eliminate one parameter in the up quark Yukawa matrix. The down quark
Yukawa matrix must then generate the mixing between the second and third
generations to get a value for $|V_{cb}|$ that agrees with experiment.
Giudice sets the parameter $n$ in the above
mass matrices to be two. We see no \'a priori reason
to suppose that this parameter must be an integer and treat it as a free
parameter.

We find the generalized one-loop
solutions (neglecting $\lambda _b$ and $\lambda _{\tau}$ in the RGEs)
\begin{eqnarray}
|V_{us}|&=&3\sqrt{{m_e}\over {m_{\mu}}}\left (
1-{25\over 2}{{m_e}\over {m_{\mu}}}
+{{4n^2}\over {9}}{{m_{\mu}}\over {m_{\tau}}}{{\eta _{\tau}}\over {\eta _{\mu}}
}\right )\;, \label{Vus} \\
|V_{cb}|&=&{{n}\over {3y}}{{m_{\mu}}\over {m_{\tau}}}
{{\eta _{\tau}}\over {\eta _{\mu}}}\left (
1-{{m_e}\over {m_{\mu}}}
+{{(n^2-3)}\over {9}}{{m_{\mu}}\over {m_{\tau}}}
{{\eta _{\tau}}\over {\eta _{\mu}}}\right )\;, \\
|V_{ub}|&=&{{y^2}\over {\eta _c}}{{m_c}\over {m_t}} \;, \\
m_u&=&y^3{{\eta _u}\over {\eta _c^2}}{{m_c^2}\over {m_t}} \;, \label{mu}\\
m_d&=&{{\eta ^{1/2}}\over {x}}{{\eta _d}\over {\eta _e}}
3m_e\left (1-8{{m_e}\over {m_{\mu}}}
+{{4n^2}\over {9}}{{m_{\mu}}\over {m_{\tau}}}
{{\eta _{\tau}}\over {\eta _{\mu}}}\right )\;, \\
m_s&=&{{\eta ^{1/2}}\over {x}}{{\eta _s}\over {\eta _{\mu}}}
{{m_{\mu}}\over {3}}
\left (1+8{{m_e}\over {m_{\mu}}}
-{{4n^2}\over {9}}{{m_{\mu}}\over {m_{\tau}}}
{{\eta _{\tau}}\over {\eta _{\mu}}}\right )\;, \\
m_b&=&y{{\eta ^{1/2}}\over {x}}{{\eta _b}\over {\eta _{\tau}}}m_{\tau}\;.
\label{mbmtaug}
\end{eqnarray}
Notice that at one-loop level to leading order in the mass heirarchy the
running $|V_{cb}|$ is related to the strange and bottom Yukawa couplings by
\begin{equation}
|V_{cb}(\mu )|=nR_{s/b}(\mu )\equiv
n{{\lambda _s(\mu )}\over {\lambda _b(\mu )}}\;.
\end{equation}

Eqs.~(\ref{Vus})-(\ref{mbmtaug})
can be compared to Eqs.~(\ref{Vusdhr})-(\ref{mbmtaudhr})
for the HRR/DHR model, except that we have retained
the highest non-leading order
corrections only for the Giudice model.
When $n=2$ the predicted value of $|V_{cb}|$ agrees well with the experimental
value. On the other hand $|V_{us}|$ is just at the lower limit of its 90\%
confidence level. The overall
situation can be improved somewhat by allowing $n$ to be
slightly larger than two.

The leading term in
Eq.~(\ref{Vus}) can be recognized as the Oakes relation\cite{Oakes}
between the Cabibbo angle and the
quark masses, $\tan \theta _c\approx \sqrt{m_d/m_s}$, supplemented by
the Yukawa unification relation $m_d/m_s=9m_e/m_{\mu}$. Notice that this
relation involving the first and second generations does not run, so the
prediction of the Cabibbo angle is insensitive to the size of the gauge and
Yukawa couplings. The
two-loop effects for the most part increase $\alpha _3$ and hence the QCD
scaling factors $\eta _q$. The influence of
two-loop contributions in the running of the Yukawas is small.

For
$\tan \beta\ltap 10$, $\lambda _b$ and $\lambda _{\tau}$ can be neglected in
the RGEs; then the relation for $m_u$ in Eq.~(\ref{mu})
implies an upper limit on $m_t$\cite{Giudice}.
However further solutions for $m_b/m_{\tau}$ are possible
with large $\tan \beta$, as can be seen
in Figure 5. In the allowed $m_b/m_{\tau}$ band at large
$\tan \beta$ the predicted value for $m_u$ from Eq.~(\ref{mu}) is still
satisfactory, since $m_t$ is in the same range as found for the small
$\tan \beta $ solutions.

The CP-violating phase is not very well constrained in the Giudice
model since the phase does not enter in the well-measured CKM elements;
in fact the phase can assume almost any non-zero value within its zero to
$2 \pi$ range.
Correspondingly CP asymmetries
to be measured in B decays are not very constrained in the model\cite{CPasy}.
In contrast, the CP-violating phase in the HRR/DHR model is almost uniquely
determined by $|V_{us}|$ and the CP-violating asymmetries are
predicted precisely.
This remain the case at the two-loop level. In the HRR/DHR scheme
the dependence on $\alpha _3(M_Z)$
cancels out in quark mass ratios, and since the constraint on the phase
arises from the first and second generation mixing angles, there is no
dependence of the phase on $\lambda _t$.

\section{Conclusion}

We have investigated unification scenarios in supersymmetric grand unified
theories using the two-loop renormalization group equations. Our primary
conclusions are the following:

(1) Given the experimentally determined values
for $\alpha _1$ and $\alpha _2$ at $M_Z$, the RGEs predict
$\alpha _3\simeq 0.111 (0.122)$ at one-loop (two-loop)
for $M_{SUSY}^{}=m_t$ and
$\alpha _3\simeq 0.106 (0.116)$ for $M_{SUSY}^{}=1$
TeV.
Including the Yukawa couplings in the two-loop evolution of the gauge couplings
decreases $\alpha _3(M_Z)$ by only a few percent. Thus the
values of $\alpha _3(M_Z)\simeq 0.12$ obtained experimentally at LEP II
are also theoretically preferred
if GUT scale thresholds effects or intermediate scales are not important.

(2) For any fixed value of $\alpha _3(M_Z)$ and $m_b$ there are just two
allowed solutions for $\tan \beta$ for a given top mass if $m_t\ltap 180$ GeV;
the larger solution
has $\tan \beta > m_t/m_b$ and the smaller solution is
$\sin \beta \simeq 0.78(m_t/150 {\rm GeV})$.
Allowing for some uncertainty in $\alpha _3(M_Z)$, $m_b$ and $M_{SUSY}^{}$,
these unique solutions
for $\tan \beta$ at given $m_t$
become a narrow range of values. For $m_t\approx 180-200$ GeV the value of
$\tan \beta$ changes rapidly with $m_t$.

(3) With $\lambda _b$, $\lambda _{\tau}$ unification we find an upper
limit $m_t\ltap 200$ GeV on the top quark mass by requiring the
successful prediction of the $m_b/m_{\tau}$ ratio; we also obtain lower limits
$m_t\gtap 150$ GeV (115 GeV) for $\alpha _3(M_Z)=0.11 (0.12)$ from
evolution constraints on $m_b$, $m_c$ and $|V_{cb}|$. These lower
limits are only mildly sensitive to $M_{SUSY}$.

(4) The effects of raising $M_{SUSY}^{}$ is to decrease
both $\alpha _G^{}$ and $M_G^{}$ and to decrease the values
of $\alpha _3(M_Z)$ that yields successful unification. Also the allowed
band for the $m_b/m_{\tau}$ ratio in the $m_t, \tan \beta $ plane
is shifted towards slightly higher top masses. This in turn
slightly reduces the prediction for $|V_{cb}|$ in models that utilize the
relation $\sqrt{\lambda _c(M_G)/\lambda _t(M_G)}=|V_{cb}(M_G)|$.

(5) In the HRR/DHR model we find an upper limit on the supersymmetry parameter
$\tan \beta \ltap 50 (60)$ for $M_{SUSY}^{}=m_t (1 {\rm TeV})$
if $\alpha _3(M_Z)\simeq 0.11$; for $\alpha _3(M_Z)=0.12$ the solutions at
large $\tan \beta $
extend into the region for which $\lambda _b(M_G)$ is non-perturbative.

(6) For the value $\alpha _3(M_Z)\simeq 0.12$ indicated by
the two-loop RGEs, the agreement of the
$|V_{cb}|$ prediction of the HRR/DHR ansatz with experiment is improved.
In fact for
$\alpha _3(M_Z)=0.12$ and $M_{SUSY}^{}=1$ TeV
the central values for $|V_{cb}|$ and the mass ratio
$m_b/m_{\tau}$ almost coincide in the $m_t,\tan \beta $ plane; see Fig. 10d.
This result
is more general than the HRR/DHR ansatz, and occurs for any model with the
GUT scale relation $|V_{cb}|=\sqrt{\lambda _c/\lambda _t}$.

(7) With $\alpha _3(M_Z)\simeq 0.12$ a large top Yukawa coupling
is needed to achieve the correct $m_b/m_{\tau}$ ratio,
and the theory is in some jeopardy of having a non-perturbative
$\lambda _t(M_G^{})$ if $m_b$ is
smaller than about 4.2 GeV.

(8) GUT unification of $\lambda _{\tau}$, $\lambda _b$ and $\lambda _t$
can be realized for $\tan \beta \gtap 50$.

(9) The predictions for the CP asymmetries in the HRR/DHR model are
largely unaffected by our two-loop analysis.

(10) We have found new solutions to the Giudice model for large $\tan \beta$.
These results require the inclusion of $\lambda _b$ and $\lambda _{\tau}$ in
the RGEs, and therefore could not be obtained in Giudice's analytic treatment
at one-loop.

\acknowledgements
One of us (VB) thanks Pierre Ramond for a discussion. One of us (MB) thanks
Greg Anderson for a discussion. We thank Tao
Han for his participation in the initial stages of this project.
This research was supported
in part by the University of Wisconsin Research Committee with funds granted by
the Wisconsin Alumni Research Foundation, in part by the U.S.~Department of
Energy under contract no.~DE-AC02-76ER00881, and in part by the Texas National
Laboratory Research Commission under grant no.~RGFY9173.  Further support was
also provided by U.S. Department of Education under Award No. P200A80214.
PO was supported in part by an NSF Graduate Fellowship.

\section{Appendix}

To consider a specific ansatz for Yukawa matrices at the GUT scale at the
two-loop level requires knowledge of the RGEs. These can be derived from
formal expressions that exist in the literature\cite{susyrge2}.
For the supersymmetric model with two Higgs doublets,
the one-\cite{susyrge1} and two-loop RGEs can be written for general Yukawa
matrices as
\begin{equation}
{{dg_i}\over {dt}}={g_i\over{16\pi^2 }}\left [b_ig_i^2+{1\over {16\pi^2 }}
\left (\sum _{j=1}^3b_{ij}g_i^2g_j^2
-\sum _{j=U,D,E}a_{ij}g_i^2
{\bf Tr}[{\bf Y_j^{}Y_j^{\dagger }}]\right )\right ] \;,
\end{equation}
with ${\bf Y_U^{}}\equiv {\bf U}$, etc.
%
%
%
\begin{eqnarray}
{{d{\bf U}}\over {dt}}={1\over {16\pi ^2}}
\Bigg[\Big [&-&\sum c_ig_i^2+3{\bf UU^{\dagger }}+{\bf DD^{\dagger }}
+{\bf Tr}[3{\bf UU^{\dagger }}]\Big ] \nonumber \\
&+&{1\over {16\pi ^2}}\Bigg (\sum \left(c_ib_i+c_i^2/2\right )g_i^4
+g_1^2g_2^2+{136\over 45}g_1^2g_3^2+8g_2^2g_3^2\nonumber \\
&&\;\;\;\;\;\;\;+({2\over 5}g_1^2+6g_2^2)
{\bf UU^{\dagger }}+{2\over 5}g_1^2{\bf DD^{\dagger }}
+({4\over 5}g_1^2+16g_3^2){\bf Tr}[{\bf UU^{\dagger }}] \nonumber \\
&&\;\;\;\;\;\;\;-9{\bf Tr}[{\bf UU^{\dagger }}{\bf UU^{\dagger }}]
-3{\bf Tr}[{\bf UU^{\dagger }}{\bf DD^{\dagger }}]
-9{\bf UU^{\dagger }}{\bf Tr}[{\bf UU^{\dagger }}]\nonumber \\
&&\;\;\;\;\;\;\;-{\bf DD^{\dagger }}{\bf Tr}[3{\bf DD^{\dagger }}
+{\bf EE^{\dagger }}]-4({\bf UU^{\dagger }})^2-2({\bf DD^{\dagger }})^2
-2{\bf UU^{\dagger }}{\bf DD^{\dagger }}\Bigg )\Bigg ]{\bf U} \nonumber \;, \\
\label{dUdt}
\end{eqnarray}
\begin{eqnarray}
{{d{\bf D}}\over {dt}}={1\over {16\pi ^2}}
\Bigg [\Big [&-&\sum c_i^{\prime}g_i^2+3{\bf DD^{\dagger }}+{\bf UU^{\dagger }}
+{\bf Tr}[3{\bf DD^{\dagger }}+{\bf EE^{\dagger }}]\Big ] \nonumber \\
&+&{1\over {16\pi ^2}}\Bigg (\sum \left(c_i^{\prime}b_i+c_i^{\prime 2}
/2\right )g_i^4
+g_1^2g_2^2+{8\over 9}g_1^2g_3^2+8g_2^2g_3^2\nonumber \\
&&\;\;\;\;\;\;\;+({4\over 5}g_1^2+6g_2^2)
{\bf DD^{\dagger }}+{4\over 5}g_1^2{\bf UU^{\dagger }}
+(-{2\over 5}g_1^2+16g_3^2){\bf Tr}[{\bf DD^{\dagger }}]
+{6\over 5}g_1^2{\bf Tr}[{\bf EE^{\dagger }}] \nonumber \\
&&\;\;\;\;\;\;\;-9{\bf Tr}[{\bf DD^{\dagger }}{\bf DD^{\dagger }}]
-3{\bf Tr}[{\bf DD^{\dagger }}{\bf UU^{\dagger }}]
-3{\bf Tr}[{\bf EE^{\dagger }}{\bf EE^{\dagger }}]
-3{\bf UU^{\dagger }}{\bf Tr}[{\bf UU^{\dagger }}] \nonumber \\
&&\;\;\;\;\;\;\;-3{\bf DD^{\dagger }}{\bf Tr}[3{\bf DD^{\dagger }}
+{\bf EE^{\dagger }}]-4({\bf DD^{\dagger }})^2-2({\bf UU^{\dagger }})^2
-2{\bf DD^{\dagger }}{\bf UU^{\dagger }}\Bigg )\Bigg ]{\bf D} \nonumber\;, \\
\label{dDdt}
\end{eqnarray}
\begin{eqnarray}
{{d{\bf E}}\over {dt}}={1\over {16\pi ^2}}
\Bigg [\Big [&-&\sum c_i^{\prime \prime}g_i^2+3{\bf EE^{\dagger }}
+{\bf Tr}[3{\bf DD^{\dagger }}+{\bf EE^{\dagger }}]\Big ] \nonumber \\
&+&{1\over {16\pi ^2}}\Bigg (\sum \left(c_i^{\prime \prime}b_i
+c_i^{\prime \prime 2}/2\right )g_i^4
+{9\over 5}g_1^2g_2^2\nonumber \\
&&\;\;\;\;\;\;\;+6g_2^2
{\bf EE^{\dagger }}
+(-{2\over 5}g_1^2+16g_3^2){\bf Tr}[{\bf DD^{\dagger }}]
+{6\over 5}g_1^2{\bf Tr}[{\bf EE^{\dagger }}] \nonumber \\
&&\;\;\;\;\;\;\;-9{\bf Tr}[{\bf DD^{\dagger }}{\bf DD^{\dagger }}]
-3{\bf Tr}[{\bf DD^{\dagger }}{\bf UU^{\dagger }}]
-3{\bf Tr}[{\bf EE^{\dagger }}{\bf EE^{\dagger }}]\nonumber \\
&&\;\;\;\;\;\;\;-3{\bf EE^{\dagger }}{\bf Tr}[3{\bf DD^{\dagger }}
+{\bf EE^{\dagger }}]-4({\bf EE^{\dagger }})^2\Bigg )\Bigg ]{\bf E}\;,
\label{dEdt}
\end{eqnarray}
where
\begin{eqnarray}
b_i&=&({33\over 5},1,-3) \;, \\
c_i&=&({13\over 15},3,{16\over 3}) \;, \\
c_i^{\prime}&=&({7\over 15},3,{16\over 3}) \;, \\
c_i^{\prime \prime}&=&({9\over 5},3,0) \;, \\
d_i&=&c_i^{\prime}-c_i^{\prime \prime} \;,
\end{eqnarray}
\begin{equation}
b_{ij} = \left( \begin{array}{c@{\quad}c@{\quad}c}
{199\over 25} & {27\over 5} & {88\over 5} \\
{9\over 5} & 25 & 24 \\ {11\over 5} & 9 & 14
\end{array} \right) \;,
\end{equation}
and
\begin{equation}
a_{ij} = \left( \begin{array}{c@{\quad}c@{\quad}c}
{26\over 5} & {14\over 5} & {18\over 5} \\
6 & 6 & 2 \\ 4 & 4 & 0
\end{array} \right) \;.
\end{equation}
%
%
%
These equations agree with those in the last paper in Ref.~\cite{susyrge2}
for the case where the Yukawa matrices are diagonal, if the following
minor corrections are made:
(1) $b_{31}$ should be decreased by a factor three;
(2) the parenthesis in the second term of $\gamma _{H_2}^{(2)}$ should come
before the $\alpha _2^2$; (3) the first term of
$\gamma _{\overline {\tau}}^{(2)}$ should have a factor
$\alpha _1^2$ instead of $\alpha _2^2$.

The two-loop RGEs for the standard model are\cite{smrge}

\begin{equation}
{{dg_i}\over {dt}}={g_i\over{16\pi^2 }}\left [b_i^{SM}g_i^2+{1\over {16\pi^2 }}
\left (\sum _{j=1}^3b_{ij}^{SM}g_i^2g_j^2
-\sum _{j=U,D,E}a_{ij}^{SM}g_i^2
{\bf Tr}[{\bf Y_j^{}Y_j^{\dagger }}]\right )\right ] \;,
\end{equation}
\begin{eqnarray}
{{d{\bf U}}\over {dt}}={1\over {16\pi ^2}}
\Bigg[\Big [&-&\sum c_i^{SM}g_i^2+{3\over 2}{\bf UU^{\dagger }}-{3\over 2}
{\bf DD^{\dagger }}
+Y_2(S)\Big ] \nonumber \\
&+&{1\over {16\pi ^2}}\Bigg ({1187\over 600}g_1^4-{23\over 4}g_2^4-108g_3^4
-{9\over 20}g_1^2g_2^2+{19\over 15}g_1^2g_3^2+9g_2^2g_3^2\nonumber \\
&&\;\;\;\;\;\;\;+\left ({223\over 80}g_1^2+{135\over 16}g_2^2+16g_3^2\right )
{\bf UU^{\dagger }}-\left ({43\over 80}g_1^2-{9\over 16}g_2^2+16g_3^2
\right ){\bf DD^{\dagger }}\nonumber \\
&&\;\;\;\;\;\;\;+{5\over 2}Y_4(S)
-2\lambda \left (3{\bf UU^{\dagger }}+{\bf DD^{\dagger }}\right )\nonumber \\
&&\;\;\;\;\;\;\;+{3\over 2}({\bf UU^{\dagger }})^2
-{\bf DD^{\dagger }}{\bf UU^{\dagger }}
-{1\over 4}{\bf UU^{\dagger }}{\bf DD^{\dagger }}
+{11\over 4}({\bf DD^{\dagger }})^2\nonumber \\
&&\;\;\;\;\;\;\;+Y_2(S)\left ({5\over 4}{\bf DD^{\dagger }}
-{9\over 4}{\bf UU^{\dagger }}\right )-\chi _4(S)+{3\over 2}\lambda ^2
\Bigg )\Bigg ]{\bf U} \nonumber\;, \\
\label{dUsmdt}
\end{eqnarray}
\begin{eqnarray}
{{d{\bf D}}\over {dt}}={1\over {16\pi ^2}}
\Bigg [\Big [&-&\sum c_i^{\prime SM}g_i^2+{3\over 2}{\bf DD^{\dagger }}
-{3\over 2}{\bf UU^{\dagger }}
+Y_2(S)\Big ] \nonumber \\
&+&{1\over {16\pi ^2}}\Bigg (-{127\over 600}g_1^4-{23\over 4}g_2^4-108g_3^4
-{27\over 20}g_1^2g_2^2+{31\over 15}g_1^2g_3^2+9g_2^2g_3^2\nonumber \\
&&\;\;\;\;\;\;\;-\left ({79\over 80}g_1^2-{9\over 16}g_2^2+16g_3^2\right )
{\bf UU^{\dagger }}+\left ({187\over 80}g_1^2+{135\over 16}g_2^2+16g_3^2
\right ){\bf DD^{\dagger }}\nonumber \\
&&\;\;\;\;\;\;\;+{5\over 2}Y_4(S)
-2\lambda \left ({\bf UU^{\dagger }}+3{\bf DD^{\dagger }}\right )\nonumber \\
&&\;\;\;\;\;\;\;+{3\over 2}({\bf DD^{\dagger }})^2
-{\bf UU^{\dagger }}{\bf DD^{\dagger }}
-{1\over 4}{\bf DD^{\dagger }}{\bf UU^{\dagger }}
+{11\over 4}({\bf UU^{\dagger }})^2\nonumber \\
&&\;\;\;\;\;\;\;+Y_2(S)\left ({5\over 4}{\bf UU^{\dagger }}
-{9\over 4}{\bf DD^{\dagger }}\right )-\chi _4(S)+{3\over 2}\lambda ^2
\Bigg )\Bigg ]{\bf D} \nonumber\;, \\
\label{dDsmdt}
\end{eqnarray}
\begin{eqnarray}
{{d{\bf E}}\over {dt}}={1\over {16\pi ^2}}
\Bigg [\Big [&-&\sum c_i^{\prime \prime SM}g_i^2+{3\over 2}{\bf EE^{\dagger }}
+Y_2(S)\Big ] \nonumber \\
&+&{1\over {16\pi ^2}}\Bigg ({1371\over 200}g_1^4-{23\over 4}g_2^4
+{27\over 20}g_1^2g_2^2\nonumber \\
&&\;\;\;\;\;\;\;+\left ({387\over 80}g_1^2+{135\over 16}g_2^2\right )
{\bf EE^{\dagger }}+{5\over 2}Y_4(S)-6\lambda {\bf EE^{\dagger }}
\nonumber \\
&&\;\;\;\;\;\;\;+{3\over 2}({\bf EE^{\dagger }})^2
-{9\over 4}Y_2(S){\bf EE^{\dagger }}-\chi _4(S)+{3\over 2}\lambda ^2
\Bigg )\Bigg ]{\bf E}\;,
\label{dEsmdt}
\end{eqnarray}
\begin{eqnarray}
{{d\lambda}\over {dt}}={1\over {16\pi ^2}}
\Bigg [&\Bigg \{&{9\over 4}\left ({3\over 25}g_1^4+{2\over 5}g_1^2g_2^2
+g_2^4\right )-\left ({9\over 5}g_1^2+9g_2^2\right )\lambda
+4Y_2(S)\lambda -4H(S)+12\lambda ^2\Bigg \} \nonumber \\
&+&{1\over {16\pi ^2}}\Bigg (-78\lambda ^3+18\left ({3\over 5}g_1^2+3g_2^2
\right )\lambda ^2+\left (-{73\over 8}g_2^4+{117\over 20}g_1^2g_2^2
+{1887\over 200}g_1^4\right )\lambda \nonumber \\
&&\;\;\;\;\;\;\;+{305\over 8}g_2^6-{867\over 120}g_1^2g_2^4
-{1677\over 200}g_1^4g_2^2-{3411\over 1000}g_1^6 \nonumber \\
&&\;\;\;\;\;\;\;-64g_3^2{\bf Tr}[({\bf UU^{\dagger }})^2
+({\bf DD^{\dagger }})^2]\nonumber \\
&&\;\;\;\;\;\;\;-{8\over 5}g_1^2{\bf Tr}[2({\bf UU^{\dagger }})^2
-({\bf DD^{\dagger }})^2+3({\bf EE^{\dagger }})^2]-{3\over 2}g_2^4Y_2(S)
+10\lambda Y_4(S)
\nonumber \\
&&\;\;\;\;\;\;\;+{3\over 5}g_1^2\left [
\left (-{57\over 10}g_1^2+21g_2^2\right ){\bf Tr}[{\bf UU^{\dagger }}]
+\left ({3\over 2}g_1^2+9g_2^2\right ){\bf Tr}[{\bf DD^{\dagger }}]\right .
\nonumber \\
&&\;\;\;\;\;\;\;\;\;\;\;\;\;\;\;\;\;\;
+\left .\left (-{15\over 2}g_1^2+11g_2^2\right ){\bf Tr}[{\bf EE^{\dagger }}]
\right ]\nonumber \\
&&\;\;\;\;\;\;\;-24\lambda ^2Y_2(S)-\lambda H(S)+6\lambda
{\bf Tr}[{\bf UU^{\dagger }}{\bf DD^{\dagger }}]\nonumber \\
&&\;\;\;\;\;\;\;+20{\bf Tr}\left [3({\bf UU^{\dagger }})^3
+3({\bf DD^{\dagger }})^3+({\bf EE^{\dagger }})^3\right ]\nonumber \\
&&\;\;\;\;\;\;\;-12{\bf Tr}\left [{\bf UU^{\dagger }}
({\bf UU^{\dagger }}+{\bf DD^{\dagger }}){\bf DD^{\dagger }}\right ]
\Bigg )\Bigg ]\;,
\label{dlamsmdt}
\end{eqnarray}
where
\begin{eqnarray}
b_i^{SM}&=&({41\over 10},-{19\over 6},-7) \;, \\
c_i^{SM}&=&({17\over 20},{9\over 4},8) \;, \\
c_i^{\prime SM}&=&({1\over 4},{9\over 4},8) \;, \\
c_i^{\prime \prime SM}&=&({9\over 4},{9\over 4},0) \;,
\end{eqnarray}
\begin{equation}
Y_2(S)={\bf Tr}[3{\bf UU^{\dagger}}+3{\bf DD^{\dagger}}+{\bf EE^{\dagger}}]\;,
\end{equation}
\begin{equation}
Y_4(S)={1\over 3}\left [3\sum c_i^{SM}g_i^2{\bf Tr}[{\bf UU^{\dagger}}]
+3\sum c_i^{\prime SM}
g_i^2{\bf Tr}[{\bf DD^{\dagger}}]+\sum c_i^{\prime \prime SM}
g_i^2{\bf Tr}[{\bf EE^{\dagger}}]\right ]\;,
\end{equation}
\begin{equation}
\chi _4(S)={9\over 4}{\bf Tr}\left [3({\bf UU^{\dagger}})^2+
3({\bf DD^{\dagger}})^2+({\bf EE^{\dagger}})^2-{2\over 3}{\bf UU^{\dagger}}
{\bf DD^{\dagger}}\right ]\;,
\end{equation}
\begin{equation}
H(S)={\bf Tr}[3({\bf UU^{\dagger}})^2+3({\bf DD^{\dagger}})^2
+({\bf EE^{\dagger}})^2]\;,
\end{equation}
\begin{equation}
b_{ij}^{SM} = \left( \begin{array}{c@{\quad}c@{\quad}c}
{199\over 50} & {27\over 10} & {44\over 5} \\
{9\over 10} & {35\over 6} & 12 \\ {11\over 10} & {9\over 2} & -26
\end{array} \right) \;,
\end{equation}
and
\begin{equation}
a_{ij}^{SM} = \left( \begin{array}{c@{\quad}c@{\quad}c}
{17\over 10} & {1\over 2} & {3\over 2} \\
{3\over 2} & {3\over 2} & {1\over 2} \\ 2 & 2 & 0
\end{array} \right) \;.
\end{equation}
These renormalization group equations are those given in the classic papers
of Machacek and Vaughn after replacing
${\bf H}\rightarrow {\bf U^{\dagger}}$,
${\bf F_D^{}}\rightarrow {\bf D^{\dagger}}$,
${\bf F_L^{}}\rightarrow {\bf E^{\dagger}}$,
and making the following corrections to
Eq.~(\ref{dlamsmdt}) mentioned
in the paper of Ford, Jack and Jones\cite{smrge}:
(1) The $\lambda g_2^2$ term in the one-loop beta function
has a coefficient 9 instead of 1.
(2) The $\lambda g_1^2g_2^2$ term in the two-loop beta function
has a coefficient $+117/20$ instead of $-117/20$.
(3) The $\lambda g_1^4$ in the two-loop beta function
has a coefficient $+1887/200$ instead of $-1119/200$.

\newpage
{\Large \center Figures}
\vskip 0.5in

\noindent Fig. 1.  Allowed GUT parameter space for $m_t=150$ GeV
with (a) $M_{SUSY}^{}=m_t$ (one-loop RGE)
(b) $M_{SUSY}^{}=m_t$ (two-loop RGE)
(c) $M_{SUSY}^{}=1$ TeV (one-loop RGE)
(d) $M_{SUSY}^{}=1$ TeV (two-loop RGE) versus the running mass scale $\mu $.
The shaded region
denotes the range of GUT coupling and mass consistent with the $1\sigma$
ranges of $\alpha _1(M_Z)$ and $\alpha _2(M_Z)$; the curves for
$\alpha _3(\mu )$ represent extrapolations from the GUT parameters. We have
omitted the contributions from Yukawa effects here which depend on
$\tan \beta$.
\vskip 0.5in

\noindent Fig. 2.  Gauge coupling unification with two-loop evolution for
(a) $M_{SUSY}^{}=m_t$
(b) $M_{SUSY}^{}=1$ TeV taking $m_t=150$ GeV and neglecting Yukawa couplings;
$\mu $ is the running mass
scale.
\vskip 0.5in

\noindent Fig. 3.  The QCD-QED scaling factors $\eta _f$ of Eq.~(\ref{etadef})
are shown for $f=s, c, b$ versus $\alpha _3(M_Z)$,
assuming running quark masses $m_f(m_f)$ of
$m_t=170$ GeV, $m_b=4.25$ GeV, $m_c=1.27$ GeV.
\vskip 0.5in

\noindent Fig. 4.  The top Yukawa coupling at the GUT scale determined
at the one-loop level is plotted versus $\alpha _3(M_Z)$ for $m_t=170$ GeV and
$m_b=4.25$ GeV.
\vskip 0.5in

\noindent Fig. 5.  Contours of constant $m_b$ in the $m_t,\tan \beta $ plane
obtained from the RGEs with
(a) $M_{SUSY}^{}=m_t$,
$\alpha _3(M_Z)=0.11$; (b) $M_{SUSY}^{}=m_t$, $\alpha _3(M_Z)=0.12$;
(c) $M_{SUSY}^{}=1$ TeV, $\alpha _3(M_Z)=0.11$; (d) $M_{SUSY}^{}=1$ TeV,
$\alpha _3(M_Z)=0.12$. The shaded band corresponds to the 90\% confidence
level range of $m_b$ from Ref.~\cite{GL} ($m_b=4.1-4.4$ GeV); the dotted curve
corresponds to $m_b=5.0$ GeV. The curves shift to higher $m_t$ values for
increasing $\alpha _3(M_Z)$ or increasing $M_{SUSY}^{}$.
\vskip 0.5in

\noindent Fig. 6.  The Yukawa couplings $\lambda _t(M_G)$
and $\lambda _b(M_G)=\lambda _{\tau}(M_G)$ at the GUT scale
with (a) $M_{SUSY}^{}=m_t$,
$\alpha _3(M_Z)=0.11$; (b) $M_{SUSY}^{}=m_t$, $\alpha _3(M_Z)=0.12$;
(c) $M_{SUSY}^{}=1$ TeV, $\alpha _3(M_Z)=0.11$; (d) $M_{SUSY}^{}=1$ TeV,
$\alpha _3(M_Z)=0.12$. The Yukawa couplings become larger for higher
$\alpha _3(M_Z)$ or higher $M_{SUSY}^{}$. The perturbative condition
$\lambda \ltap 3.3$ from Eq.~(\ref{tlim}) is satisfied except for the lowest
$b$ mass value $m_b=4.1$ GeV for
$\alpha _3(M_Z)=0.12$. The solid dots denote
$\lambda _{\tau}=\lambda _b=\lambda _t$ unification.
\vskip 0.5in

\noindent Fig. 7.  Two-loop evolution
of the Yukawa couplings (a) $\lambda _t(\mu )$
(b) $\lambda _b(\mu )$, $\lambda _{\tau}(\mu )$
from low energies to the GUT scale
for the case $\alpha _3(M_Z)=0.12$ and $M_{SUSY}^{}=1$ TeV. We take
$\tan \beta=20$ and the values of $m_t=198,\; 197,\; 196,\; 181$ GeV
specified by the
$m_b=4.1,\; 4.25,\;4.4,\; 5.0$ GeV contours in Fig 5d.
\vskip 0.5in

\noindent Fig. 8.  Two-loop evolution of the quark Yukawa ratio
$R_{c/t}\equiv \lambda _c/\lambda _t$ and the CKM matrix element $|V_{cb}|$
for (a) $M_{SUSY}^{}=m_t$ and (b) $M_{SUSY}^{}=1$ TeV.
We have taken $\alpha _3=0.11$, $\tan \beta =5$ and have chosen
the top and bottom quark masses such that
$\sqrt{R_{c/t}(M_G)}=|V_{cb}(M_G)|$ and $m_c=1.27$ GeV:
(a) $|V_{cb}(m_t)|=0.054$, $m_t=180$ GeV, $m_b=4.33$ GeV;
(b) $|V_{cb}(m_t)|=0.050$, $m_t=189$ GeV, $m_b=4.14$ GeV.
\vskip 0.5in

\noindent Fig. 9.  Contours for constant $|V_{cb}|$ at fixed $m_c=1.27$ GeV in
the $m_t,\tan \beta $ plane obtained from the RGEs with
(a) $M_{SUSY}^{}=m_t$, $\alpha _3(M_Z)=0.11$;
(b) $M_{SUSY}^{}=1$ TeV, $\alpha _3(M_Z)=0.12$.
\vskip 0.5in

\noindent Fig. 10.  Comparison of contours for constant $|V_{cb}|$ and constant
$m_b$ in the $m_t,\tan \beta $ plane from the RGEs, taking $m_c=1.27$ GeV, for
(a) $M_{SUSY}^{}=m_t$,
$\alpha _3(M_Z)=0.11$; (b) $M_{SUSY}^{}=m_t$, $\alpha _3(M_Z)=0.12$;
(c) $M_{SUSY}^{}=1$ TeV, $\alpha _3(M_Z)=0.11$; (d) $M_{SUSY}^{}=1$ TeV,
$\alpha _3(M_Z)=0.12$. The shaded band indicates the region where the 90\%
confidence limit is satisfied for $m_b$. The right-most contours are
discontinued when $\lambda _t(M_G)$ exceeds 6.
\vskip 0.5in

\noindent Fig. 11.  Comparison of contours for constant $m_c$ and constant
$m_b$ in the $m_t,\tan \beta $ plane from the RGEs, taking $|V_{cb}|$ equal to
its upper limit 0.54, for
(a) $M_{SUSY}^{}=m_t$,
$\alpha _3(M_Z)=0.11$; (b) $M_{SUSY}^{}=m_t$, $\alpha _3(M_Z)=0.12$;
(c) $M_{SUSY}^{}=1$ TeV, $\alpha _3(M_Z)=0.11$; (d) $M_{SUSY}^{}=1$ TeV,
$\alpha _3(M_Z)=0.12$. The shaded band indicates the region where 90\%
confidence limits are satisfied for all three constraints: $m_b$, $m_c$ and
$|V_{cb}|$. An {\bf X} marks the lower limit
of this shaded band and corresponds
to the values in Table 5.
\vskip 0.5in


\begin{references}
\bibitem{translate} In the notation of Ref.~\cite{DHR}
these parameters are $y ^{-1}=
(\lambda _t(M_G^{})/\lambda _t)^{1/6}\eta ^{1/12}$. In the
notation of Ref.~\cite{Giudice}, $y = \xi$ and  $x=(\gamma _D/\gamma _E)^{-1}
\eta ^{1/2}$.
\end{references}
\end{document}